\documentclass[prl,twocolumn,superscriptaddress]{revtex4-2}
\usepackage{amsmath,amssymb,mathrsfs}
\usepackage{dsfont}
\usepackage{graphicx}
\usepackage{colortbl}
\usepackage{braket}
\usepackage{bm}
\usepackage{amsfonts}
\usepackage{tabularx}
\usepackage{array}
\usepackage{multirow}
\usepackage{booktabs}
\usepackage[normalem]{ulem}
\usepackage[export]{adjustbox}
\usepackage[colorlinks,plainpages=false,linkcolor=blue,urlcolor=blue,citecolor=blue,pdfpagemode=UseNone,pdfstartview=FitBH]{hyperref}

\begin{document}
\title{Spontaneous Parity Breaking in Quantum Antiferromagnets on the Triangular Lattice}

\author{Songtai Lv}
\affiliation{Key Laboratory of Polar Materials and Devices (MOE), School of Physics, East China Normal University, Shanghai 200241, China}

\author{Yuchen Meng}
\affiliation {Key Laboratory of Polar Materials and Devices (MOE), School of Physics, East China Normal University, Shanghai 200241, China}

\author{Haiyuan Zou}
\altaffiliation{hyzou@phy.ecnu.edu.cn}
\affiliation{Key Laboratory of Polar Materials and Devices (MOE), School of Physics, East China Normal University, Shanghai 200241, China}

\begin{abstract}
Frustration on the triangular lattice has long been a source of intriguing and often debated phases in many-body systems. Although symmetry analysis has been employed, the role of the seemingly trivial parity symmetry has received little attention. In this work, we show that phases induced by frustration are systematically shaped by an implicit rule{-}of{-}thumb associated with spontaneous parity breaking {in weak longitudinal field}. This principle enables us to anticipate and rationalize the regimes and conditions under which nontrivial phases emerge.
For the spin-$S$ antiferromagnetic XXZ model, we demonstrate that a controversial parity-broken phase appears at intermediate values of $S$. In bilayer systems, enhanced frustration leads to additional phases, such as supersolids, whose properties can be classified by their characteristic parity features. Benefiting from our improved tensor network contraction techniques, we confirm these results through large-scale tensor-network calculations.
This study offers an alternative viewpoint and a systematic approach for examining the interplay between spin, symmetry, and frustration in many-body systems.
\end{abstract}

\maketitle

{\it Introduction.} 
Geometrically frustrated systems, with the triangular lattice as a paradigmatic example, host a wide variety of exotic phases and emergent phenomena~ \cite{Balents2010,Powell2011,Starykh2015,Schmidt2017,Inosov2019,Khatua2023}, including but not limited to supersolids, quantum spin liquids, and unconventional critical behaviors~\cite{Wessel2005,Heidarian2005,Melko2005,Boninsegni2005,Shimizu2003,Xu2012,Liu2018,Gallegos2025, Rau2018,Chen2024,Chen2024prr}. When an external magnetic field is applied, additional richness arises in the form of magnetization plateaus~\cite{Miyashita1986,Chubukov1991, Nojiri1988,Inami1996,Fortune2009}, analogous to Hall plateaus in electronic systems. 
Even in the absence of explicitly introduced symmetry-breaking terms such as Dzyaloshinskii–Moriya interactions, quantum fluctuations alone can stabilize numerous phases in the vicinity of these 
plateaus~\cite{Yoshikawa2004,Starykh2014,Yamamoto2014,Sellmann2015,Yamamoto2021}. Some of these phases remain under debate; however, such controversies are themselves informative, as they motivate the search for deeper organizing principles underlying phase competition in frustrated systems.

In recent years, experimental progress on triangular-lattice magnets has advanced rapidly, while a number of exotic phases continue to defy clear understanding~\cite{Inosov2019,Fortune2021prb,Zhang2024,YuYu2026}.
Various theoretical and numerical approaches, including cluster methods~\cite{Zeng1998,Yamamoto2009}, density-matrix renormalization group (DMRG)~\cite{Yoshikawa2004,White2007,Sellmann2015}, and tensor-network methods~\cite{Harada2012,Xie2014,Li2022,Sherman2023prb,Chi2022prl,Chi2024prbl}, have been employed to investigate these phase competitions; however, the very richness of phases in triangular-lattice systems inevitably leads to pronounced method dependence: different numerical approaches may yield conflicting conclusions regarding the existence or stability region of a given phase, even for states with well-established ordering patterns, let alone quantum spin liquids. This situation underscores the need for more powerful and reliable numerical methods, which must incorporate physically motivated entanglement structures~\cite{Zou2023SCPMA} and respect fundamental theoretical constraints intrinsic to frustrated systems.

\begin{figure}[t]
\includegraphics[width=0.4\textwidth]{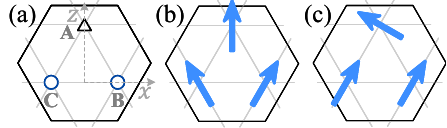}
\caption{Illustrations of representative phases. (a) $A$, $B$, and $C$ sites are coplanar in the $x$–$z$ plane. (b,c) show the spin configurations of the $\pi$-coplanar and the V phase, respectively.
}
\label{fig.fv}
\end{figure}

For this reason, the symmetries of the system are usually taken into account~\cite{Starykh2015}; however, parity, which is crucial in the study of topology~\cite{Read2000}, has rarely been employed as a guiding principle for analyzing its impact on frustrated quantum many-body systems.
As two counterintuitive examples in antiferromagnetic (AF) spin{-$S$} models on the triangular lattice~{(Fig.~\ref{fig.fv})}, 
{consider the $\pi$-coplanar phase~\mbox{\cite{Yamamoto2014,Yamamoto2017,Nikuni1995}} (also referred to as the $\Psi$~\cite{Starykh2014} or Fan phase~\cite{Maryasin2013}), which appears mirror-symmetric but actually breaks parity inversion ($B\leftrightarrow C$) about site $A$. In contrast, the mirror-asymmetric V phase preserves parity. All other phases can be similarly characterized~\cite{supp}, 
with the parity breaking centered at $A$ captured by the order parameter $P_A=|S_B-S_C|$, and the overall parity breaking described by the order parameter $P=\left(|S_B-S_C|+|S_C-S_A|+|S_A-S_B|\right)/3$, defined as the average over the different sublattice sites.
This observation raises a question}: are there universal parity principles that can account for the existence and stability of these phases? Moreover, the emergence of phases with classical analog configurations is typically driven by quantum order-by-disorder effects, whose strength depends on the spin $S$. Numerical results such as tensor networks at larger $S$ are therefore crucial. However, the rapidly increasing computational complexity has so far limited systematic comparisons between numerical results and theoretical expectations.

\begin{figure*}[t]
\includegraphics[width=0.85\textwidth]{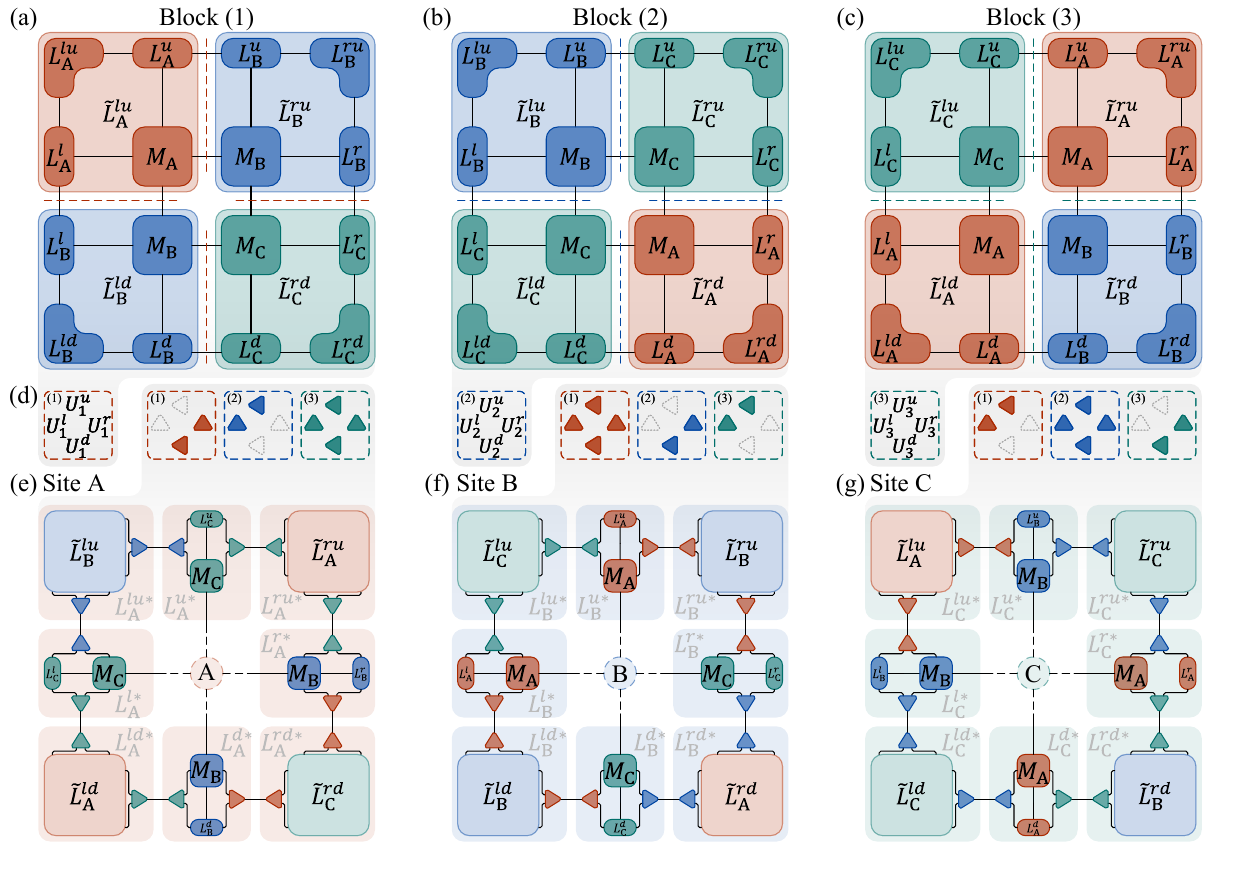}
\caption{
CTMRG procedure for the triangular lattice. (a–c) Three blocks arising from the triangular-lattice periodicity, each composed of local tensors $M_i$ ($i$ = $A$, $B$, $C$) together with their associated edge tensors $L_i^\mu$ ($\mu =$ $l$, $d$, $r$, $u$) and corner tensors $L_i^{\mu\nu}$ ($\mu\nu =$ $lu$, $ld$, $rd$, $ru$). (d) Isometries $U_n^\mu$ ($n$ = 1, 2, 3) obtained from singular-value decompositions at the dashed cuts in (a-c) are shown. Solid triangles indicate the position of $U_n^\mu$ for the next step. (e–g) Update of the environment tensors through contractions with $U_n^\mu$ and $U_n^{\mu\dagger}$, yielding the renormalized $L_i^{\mu *}$ and $L_i^{\mu\nu *}$.
}
\label{fig.CTMRG}
\end{figure*}

Motivated by these considerations, as well as by the interpretation of the celebrated Lee–Yang result~\cite{LeeYang1956parity} on parity violation in weak interactions provided by the effective $V$–$A$ theory~\mbox{\cite{Feynman1958,Sudarshan1958}}, we are led to analogously analyze the effective Hamiltonians describing the ground states of frustrated systems (with detail in the End Matter), which implies a rule-of-thumb parity principle that applies to the AF triangular-lattice XXZ (TLXXZ) model for arbitrary $S$: in frustration-induced spin-ordered phases, parity is spontaneously broken in weak longitudinal magnetic field, while increasing the field tends to restore parity symmetry. As a corollary, paramagnetic phases (e.g., valence-bond-solid states) stabilized by additional frustration (e.g., in bilayer systems) are expected to preserve parity, whereas tuning the field which compete with frustration may lead to sequences in which parity is first broken and subsequently restored.

To substantiate these ideas, we develop an improved and accelerated corner-transfer-matrix renormalization group (CTMRG) contraction scheme, enabling large-scale tensor-network simulations for different $S$ as well as bilayer geometries. Our results confirm the proposed parity principle, clarify the long-debated existence and parameter regime of the {$\pi$-coplanar} phase at large $S$, and characterize the parity properties of supersolid phases in bilayer systems. By resolving several long-standing and controversial theoretical issues, our findings also provide valuable insight for interpreting experimental observations in triangular-lattice magnets.

{\it TLXXZ model.} 
We first consider the two-dimensional TLXXZ model, described by the Hamiltonian,
\begin{equation}
    H = 
       J_{ab} \sum_{\braket{i,j}_{ab}} (S_i^x S_j^x + S_i^y S_j^y + \eta S_i^z S_j^z) 
      - h_z \sum_{i} S_i^z,
    \label{eq.H}
\end{equation}
where $J_{ab}$ denotes the nearest-neighbor interaction. The operator $S_i$ represents the spin at site $i$, the anisotropy parameter $\eta \le 1$ is considered, and $h_z$ is the longitudinal magnetic field.
For $S=1/2$, the phase diagram in the plane of $h_z$ and $\eta$ is widely studied~\cite{Yamamoto2014,Sellmann2015}. It includes phases such as the umbrella phase, the up–up–down (UUD) phase, the Y phase, and the V phase. The existence and location of the {$\pi$-coplanar} phase, however, remain controversial: cluster-based approaches place it between the ferromagnetic (FM), V, and umbrella phases~\cite{Yamamoto2014}, whereas DMRG calculations have not found evidence for its stability~\cite{Sellmann2015}.

We also consider the bilayer TLXXZ model, for which the Hamiltonian contains an additional interlayer term
\begin{equation}
H'=J_{c} \sum_{\braket{i,j}_c}    \vec{S}_i \cdot \vec{S}_j+ J_{d} \sum_{\langle\!\braket{i,j}\!\rangle}\vec{S}_i \cdot \vec{S}_j,
\end{equation}
where $J_c$ denotes the coupling along the $c$-axis between the interlayer pairs $\langle i,j\rangle_c$. The coupling $J_d$ represents the interlayer next-nearest-neighbor interaction between pairs $\langle\!\langle i,j\rangle\!\rangle$. In the absence of a magnetic field, sufficiently large $J_c$ drives the bilayer system into a dimerized phase with interlayer spin singlets. Upon increasing the longitudinal field, the system successively exhibits the 1/3 and 2/3 magnetization plateaus before saturation, as well as additional supersolid phases and Bose–Einstein–condensed (BEC) phases~\cite{Yamamoto2013a}.

\begin{figure}[t]
\includegraphics[width=0.5\textwidth]{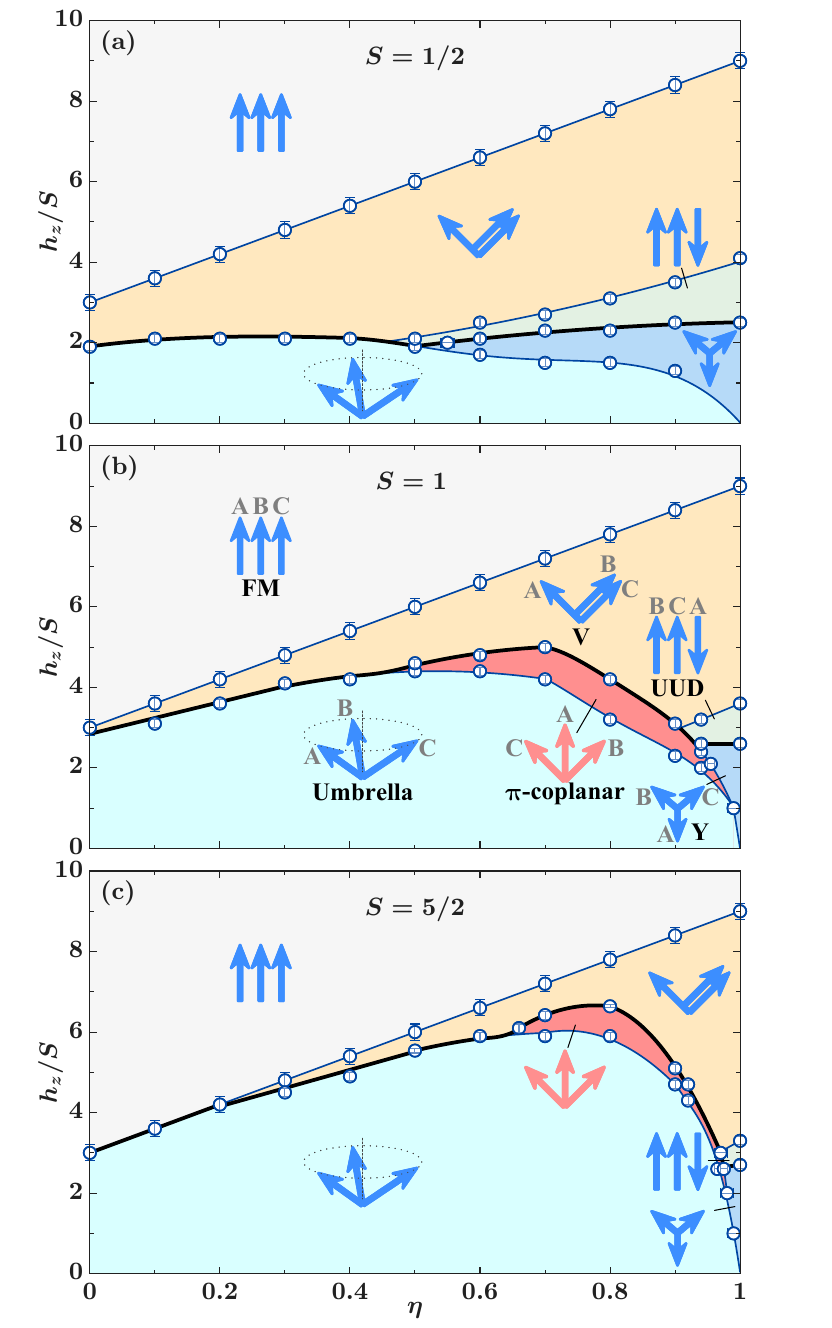}
\caption{Ground state phase diagrams of the monolayer TLXXZ model for $S =$ 1/2, 1, and 5/2, shown in panels (a-c), respectively. Circles with error bars are the calculated boundaries, while solid curves are guides to the eye. Thick black lines denote parity-breaking phase transitions. Arrows represent spin configurations (FM, Umbrella, V, Y, UUD, and {$\pi$-coplanar}) on the $A$, $B$, and $C$ sublattices.
}
\label{fig.phaseDiagram}
\end{figure}

{\it CTMRG procedure for the TLXXZ model.} 
Following Ref.~\cite{Li2022}, we construct a projected entangled simplex state (PESS) ansatz~\cite{Xie2014} using a unit cell consisting of three local tensors and three simplex tensors, and evolve the state until convergence. By contracting each local tensor with the associated simplex tensors and tracing over the physical indices, we obtain three rank-four tensors $M_i$, ($i=$ $A$,$B$,$C$), which serve as the building blocks for two-dimensional coarse graining. This contraction is carried out using the CTMRG method. Originating from a DMRG algorithm~\cite{Nishino1996}, CTMRG has undergone substantial developments alongside tensor-network techniques~\cite{Corboz2010,Orus2009,Lukin2023,Nyckees2023,Zhang2025,Chen2025} and has become a key tool for tensor-network simulations in the thermodynamic limit.

Here, we propose a more highly parallelized CTMRG scheme that significantly improves computational efficiency {in the case of unit cells containing inequivalent tensors $M_i$. Conventional CTMRG methods typically either assign a common set of environments to a large block composed of $M_i$~\cite{Corboz2010}, which necessitates multiple singular value decomposition (SVD) cut steps along a single direction to construct the dimension-reducing isometry $U$, or obtain $U$ via repeated directional cuts during multiple evolution sweeps~\cite{Chen2025}. In contrast, our method completely separates the construction of $U$ from the evolution process. 
We assign an independent set of environments to each $M_i$, and precompute all required 
$U$s from all possible blocks prior to evolution. These $U$s are then placed appropriately to evolve the environments, avoiding both multiple cuts along one direction and the iterative evolution-and-cut along four directions.} 
 
Specifically, taking the case of three tensors as an example, the environment of each tensor $M_i$ is constructed from four edge tensors and four corner tensors {[marked by the same color in Fig.~\ref{fig.CTMRG}(a-c)]}. By considering configurations composed of four $M_i$ tensors, all three possible {Blocks} are included. For each {Block}, matrices are formed by cutting the network along the four lattice directions (indicated by dashed lines), thereby encoding the full environmental information of $M_i$ in the up, down, left, and right directions. SVD of these matrices yields twelve isometries $U_i$ {[Fig.~\ref{fig.CTMRG}(d)]}, which are then simultaneously inserted on both sides of all environment tensors according to the lattice geometry, realizing a one-step evolution of the entire environment~{[Fig.~\ref{fig.CTMRG}(e-g)]}. After convergence, physical observables 
can be readily evaluated.
Applying this modified method to the triangular-lattice antiferromagnetic Heisenberg model at zero field yields a ground-state energy consistent with those obtained from other CTMRG-based methods~\cite{Li2022,supp}.

{\it Phase diagrams of the monolayer TLXXZ model.} 
Using the above-developed method, we perform large-scale simulations of the TLXXZ model for $S=$1/2, 1, and 5/2, and obtain comprehensive ground-state phase diagrams in the {$\eta$–$h_z$} plane (Fig.~\ref{fig.phaseDiagram}). For the $S=1/2$ case, aside from some minor differences in details, our results for the coplanar V, UUD, and Y phases, as well as the noncoplanar umbrella phase, are in agreement with previous studies.~\cite{Yamamoto2014,Sellmann2015}.
Notably, we do not find any evidence for the {$\pi$-coplanar} phase, in agreement with DMRG results~\cite{Sellmann2015}. 
{This absence is in agreement with the parity rule-of-thumb discussed above.}
At zero field, frustration stabilizes the $120^\circ$ ground state and breaks parity symmetry. 
As $h_z$ increases, parity-broken phases (Y and umbrella) give way to parity-preserving phases (UUD and V). Further increasing $h_z$ continues to protect parity symmetry, making it difficult for a parity-breaking {$\pi$-coplanar} phase to emerge between the FM and V phases. 

In fact, 
{the parity rule-of thumb implies}
that the {$\pi$-coplanar} phase, if present, must appear before the parity-preserving V phase as $h_z$ is increased.
Our results for larger spins, $S=$1 and 5/2, are fully consistent with this picture. With increasing $S$, the umbrella phase shifts to higher fields and eventually connects directly to the FM phase in the classical limit, 
whereas all other phases are squeezed toward $\eta=1$ and become degenerate there.
{Beyond the spin configurations, the distinctions between the phases can be further captured by order parameters related to chirality~\cite{Yoshikawa2004}, nematicity~\cite{Xu2012}, and parity. In particular, the parity order parameters provide an intuitive and quantitative picture of spontaneous symmetry breaking. An illustrative example for $S=1$ and $\eta=0.8$ is given in the End Matter.}

\begin{figure}[t]
\includegraphics[width=0.5\textwidth]{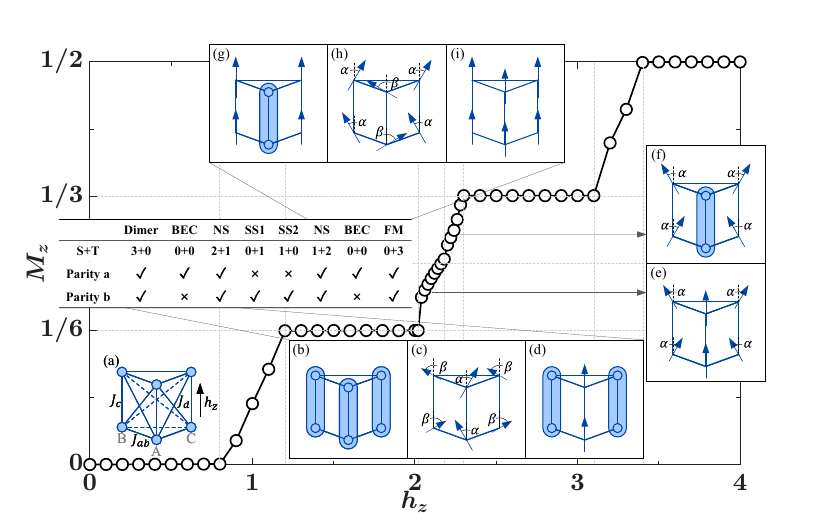}
\caption{
Magnetization $M_z$ of the bilayer TLXXZ model with varied $h_z$. Inset (a) illustrates the lattice geometry and couplings, and insets (b–i) show representative spin configurations of each phase. The table lists the numbers of singlets (S) and triplets (T), as well as whether parity a and parity b are preserved.
}
\label{fig.Mz}
\end{figure}

{\it Dimer breaking in the bilayer TLXXZ model.} 
Our approach can be straightforwardly extended to bilayer systems, where additional couplings give rise to competition among a much richer set of exotic phases. Here we focus on the $S=1/2$ isotropic case ($\eta = 1$) and choose a parameter regime with dominant interlayer coupling ($J_{ab}$ = 0.425, $J_c$ = 1, $J_d$ = 0.325) to stabilize interlayer dimerization. As $h_z$ is increased, we explicitly confirm the emergence of the 1/3 and 2/3 magnetization plateaus [or normal solid (NS) states]~\cite{Yamamoto2013a}, corresponding respectively to configurations with two remaining singlets and with one remaining singlet per unit cell during the progressive destruction of dimers (Fig.~\ref{fig.Mz}). We identify all spin configurations along this process, and find that the spins are coplanar in all these phases. Between the plateaus and the fully dimerized or fully polarized limits, the system enters BEC phases in which each individual layer realizes a V-type configuration. Between the two plateaus, two supersolid phases (SS1 and SS2) appear and exhibit {$\pi$-coplanar}-like structure in each layer~\cite{supp}.

We interpret this evolution again from the perspective of spontaneous parity breaking. Unlike the single-layer case, where the $h_z$ competes mainly with intralayer antiferromagnetism, here the field must additionally compete with multiple interlayer couplings. First, at $h_z = 0$, the strong AF interlayer coupling ($J_c$) favors dimer formation and thereby restores parity symmetry. When $J_{ab} = J_d$, the competition between these two couplings effectively cancels interactions among the three dimers, such that increasing $h_z$ drives the system directly from the fully dimerized state through the 1/3 and 2/3 plateaus to the fully polarized state. In this case, dimers are replaced by triplets one by one without breaking parity. The BEC and supersolid phases appear only when $J_{ab} \neq J_d$, where directional correlations among dimers become energetically favored and intermediate phases are stabilized between the plateaus.

In a bilayer structure, two distinct parities can be considered. The first, denoted parity a, treats each interlayer dimer as an effective unit and is thus analogous to the parity symmetry of the single-layer system. The second, parity b, resolves the bilayer structure explicitly, with the parity center chosen at the midpoint between the two layers at site A. For parity a, the transitions from the three-dimer state to the 1/3 plateau, and from the 2/3 plateau to FM phase, involve only modifications of the configuration at site A; the exchange of B and C sublattices under parity remains unaffected. Therefore, the intermediate BEC phases preserve parity a. In contrast, the supersolid phases necessarily involve choosing either the B or C sublattice on which a dimer is removed, thereby breaking parity a. 
For parity b, which resolves the detailed bilayer structure, the situation depends more sensitively on the competition between $J_{ab}$ and $J_d$. During the transition from the three-dimer state to the 1/3 plateau, the BEC phase involves the breaking of dimers on the B and C sites as well, and is accompanied by the breaking of parity b. By contrast, in the two supersolid phases, we find that parity b is remarkably preserved. This reveals a more subtle and distinctive internal parity structure of supersolid phases in bilayer systems, which is absent in their single-layer counterparts.

{~\it Summary and outlook. }
In this work, by systematically investigating the AF TLXXZ model for $S =$ 1/2, 1, and 5/2 using an improved CTMRG scheme, we uncover a previously overlooked rule-of-thumb governing parity breaking, as well as the impact of frustration on parity. With increasing external longitudinal field, parity generally exhibits a tendency toward symmetry restoration. The long-debated {$\pi$-coplanar} phase is found to emerge for 
$S \ge 1$ and at fields below the supersolid phase, providing an important reference for further studies of the superfluid-to-supersolid transition. In bilayer systems, the presence of additional frustration leads to phases such as supersolids with much richer parity structures.
Our results have direct implications for a broad class of triangular-lattice experiments. For instance, in spin-5/2 materials, the extent of the 1/3 magnetization plateau varies significantly among different compounds, with some exhibiting extremely narrow plateau regions~\cite{Smirnov2007,Tian2014,Yu2026exp}. 
Building on our results, the inclusion of further frustration effects, such as next-nearest-neighbor interactions~\cite{Gallegos2025} and transverse fields~\cite{Yamamoto2015,Yamamoto2019}, may provide a natural explanation for additional exotic plateaus, including the 1/2 and 5/6 plateaus observed experimentally~\cite{YuYu2026}. More broadly, our analysis of the interplay between parity and frustration, {and especially the intuitive and quantitative description of parity breaking,} can be extended to other frustrated lattices.
The modified CTMRG method presented here also offers a powerful and efficient tool for the study of quantum many-body physics.

{~\it Acknowledgments—}We thank Jie Ma, Tian Shang, Deyan Sun, Zhiyuan Xie, Daisuke Yamamoto, Rong Yu, and Weiqiang Yu for helpful discussions. 
This work is supported by the National Natural Science Foundation of China (Grant No. 12274126). 

%

\section{End Matter}

\makeatletter
\setcounter{figure}{0}
\renewcommand{\thefigure}{A\arabic{figure}}

{\bf Appendix A: Connection to the Lee–Yang parity violation in weak interactions.}
Historically, the $V$-$A$ theory~\mbox{\cite{Feynman1958,Sudarshan1958}} described explicit parity violation in weak interactions~\cite{LeeYang1956parity} by combining the vector $V$ and axial-vector (pseudovector) $A$, which transform oppositely under parity, into a chiral operator. In contrast, the parity breaking in certain weak-field ground states of the TLXXZ model is {\bf spontaneous}, as the Hamiltonian remains parity invariant. Although only pseudovector appears in the spin model, we can still draw inspiration from the $V$-$A$ theory and, at the mean-field level, decompose spin expectation values
$\langle S^\alpha_B\rangle$ and $\langle S^\alpha_C\rangle$ ($\alpha=x$, $y$, or $z$) into a combination of $\langle S^{\alpha+}_{BC}\rangle$ and $\langle S^{\alpha-}_{BC}\rangle$, where $\langle S^{\alpha+}_{BC}\rangle=(\langle S^\alpha_B\rangle+\langle S^\alpha_C\rangle)/2$ and $\langle S^{\alpha-}_{BC}\rangle=(\langle S^\alpha_B\rangle-\langle S^\alpha_C\rangle)/2$, with the latter serving as a parity order parameter. In this way, the original Hamiltonian Eq.~(\ref{eq.H}) within one unit cell can be rewritten as the following effective model:
\begin{equation}
  \tilde{H} = \sum_{\alpha}\left[\tilde{H}^\alpha-\xi^\alpha\langle S^{\alpha-}_{BC}\rangle(S^{\alpha}_B-S^{\alpha}_C)+\langle S^{\alpha-}_{BC}\rangle^2\right],
    \label{eq.Heff}
\end{equation}
{where} 
\begin{align}
    \tilde{H}^\alpha= &\left(\langle S^\alpha_A\rangle+\langle S^{\alpha+}_{BC}\rangle\right)\left(S^{\alpha}_B+S^{\alpha}_C\right)-\langle S^{\alpha+}_{BC}\rangle^2 \\ \notag
    & +2\langle S^{\alpha+}_{BC}\rangle S^{\alpha}_A-2\langle S^{\alpha+}_{BC}\rangle \langle S^{\alpha}_{A}\rangle
\end{align}
%
%
{for $\alpha=x$ or $y$,}
\begin{align}
    \tilde{H}^z= &\left[\eta\left(\langle S^z_A\rangle+\langle S^{z+}_{BC}\rangle\right)-h_z/3\right]\left(S^{z}_B+S^z_C\right)-\eta\langle S^{z+}_{BC}\rangle^2 \\ \notag
    & + \left(2\eta\langle S^{z+}_{BC}\rangle-h_z/3\right) S^z_A-2\eta\langle S^{z+}_{BC}\rangle\langle S^{\alpha}_{A}\rangle,
\end{align}
{and $\vec{\xi}=(1,1,\eta)$. Note that, if one adopts the slave-fermion representation, $\langle S^{\alpha-}_{BC}\rangle$ can be expressed as a bilinear form of fermionic quasiparticle operators 
$c$ and $c^\dagger$. With the mean-field order parameter $\phi\equiv\langle c\rangle=\langle c^\dagger\rangle$, the effective Hamiltonian Eq.~(\ref{eq.Heff}) naturally takes the Landau form $-a\phi^2+b\phi^4$, which describes spontaneous symmetry breaking. Furthermore, the $\langle S^{\alpha-}_{BC}\rangle$ terms originate solely from the interaction $S_BS_C$, whereas $\langle S^{\alpha+}_{BC}\rangle$ terms also receive contributions from the coupling with site $A$. The opposite signs of $\langle S^{\alpha-}_{BC}\rangle^2$ and $\langle S^{\alpha+}_{BC}\rangle^2$ lead to their different responses to the field $h_z$. 
As $h_z$ is lowered from the fully symmetric FM state ($A$, $B$, $C$ equivalent) to the coplanar phase (in the $x$-$z$ plane) and then to the non-coplanar phase, the deformation of this symmetric ground state makes the choice of site $A$ arbitrary. The positive $\langle S^{\alpha-}_{BC}\rangle^2$ term, being decoupled from $A$, tends to keep $\langle S^{\alpha-}_{BC}\rangle=0$, while the negative $\langle S^{\alpha+}_{BC}\rangle^2$ term, coupled to variations of $A$, favors a finite $\langle S^{\alpha+}_{BC}\rangle$, thus modifying $\tilde{H}^x$. As $h_z$ decreases further, the $\tilde{H}^z$ energy increases, driving the system to spontaneously generate a finite $\langle S^{\alpha-}_{BC}\rangle$ to reduce the energy. When $h_z$ becomes low enough for $\tilde{H}^y$ to contribute, a non-coplanar phase appears.}

\begin{figure}[htbp]
    \includegraphics[width=0.45\textwidth]{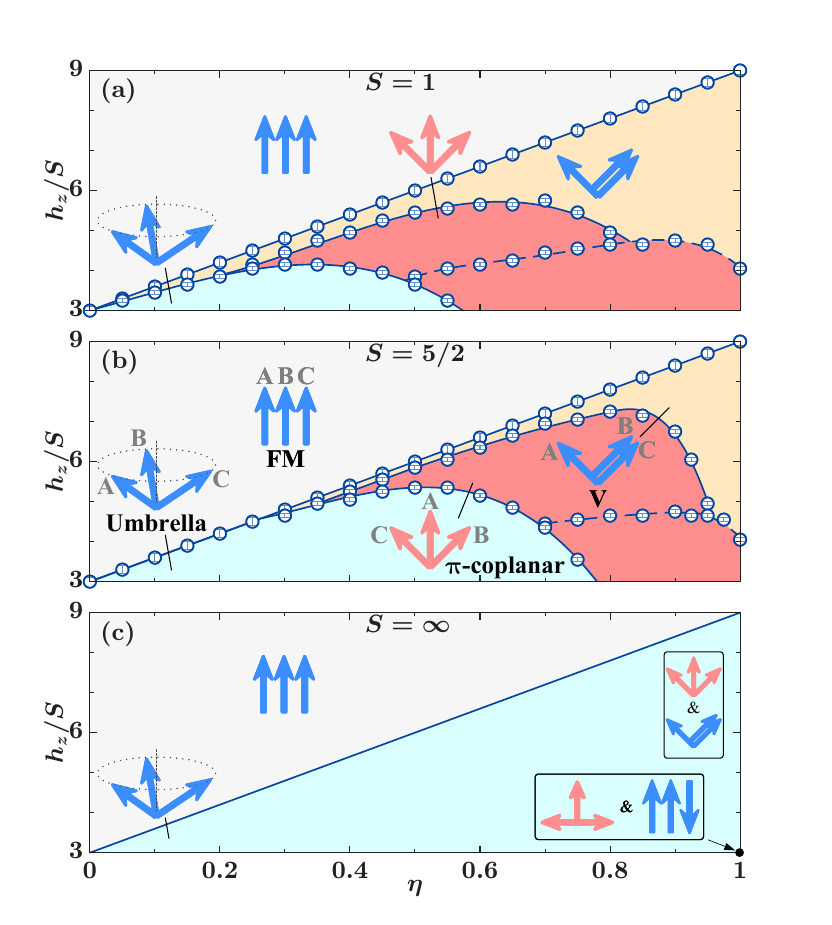}%
    \caption{Mean Field phase diagrams of the monolayer TLXXZ model, with notations following Fig.~\ref{fig.phaseDiagram}. (a,b) LSW results for $S=1$ and 5/2; the V phase is unstable below the dashed line. (c) Exact classical results. At $\eta=1$, the $\pi$-coplanar and V phases are degenerate with the umbrella phase. At $h_z/S=3$, the V phase becomes UUD phase, and the spin on $B$ (or $C$) in the $\pi$-coplanar phase is perpendicular to the spin on $A$.}
    \label{fig:figA1}
\end{figure}

{To further illustrate the above analogy, we perform linear spin-wave (LSW) calculations for finite $S$ as well as exact analyses in the classical limit ($S\rightarrow\infty$)~\mbox{\cite{Colpa1978,supp}}. 
All results for general $S$ follow the rule-of-thumb that lowering $h_z$ from the FM state leads to spontaneous parity breaking, either through the V, $\pi$-coplanar, and umbrella phases for finite $S$, or directly to the umbrella phase in the $S\rightarrow\infty$ limit.
In the LSW calculations, we construct the ansatz for each phase and determine the ground state by comparing the corresponding energies. For $S=1$ and 5/2 [Fig.~\ref{fig:figA1}(a,b)], the results are consistent with our tensor-network numerical calculations.
As $S$ increases, both the $\pi$-coplanar and V phases are compressed toward $\eta=1$ and become degenerate with the umbrella phase in the classical limit [Fig.~\ref{fig:figA1}(c)]. 
For example, the exact classical ground-state energies of the umbrella and $\pi$-coplanar phases are $E_{U}=-(h^2_z+18\eta+9)/[6(1+2\eta)]$ and $E_\pi=-[h^2_z+3h_z(1-\eta)+9(\eta^2+\eta+1)]/[9(1+\eta)]$, respectively, with $E_{U}\le E_{\pi}$, and they coincide at $\eta=1$. }

\begin{figure}[htbp]
    \includegraphics[width=0.49\textwidth]{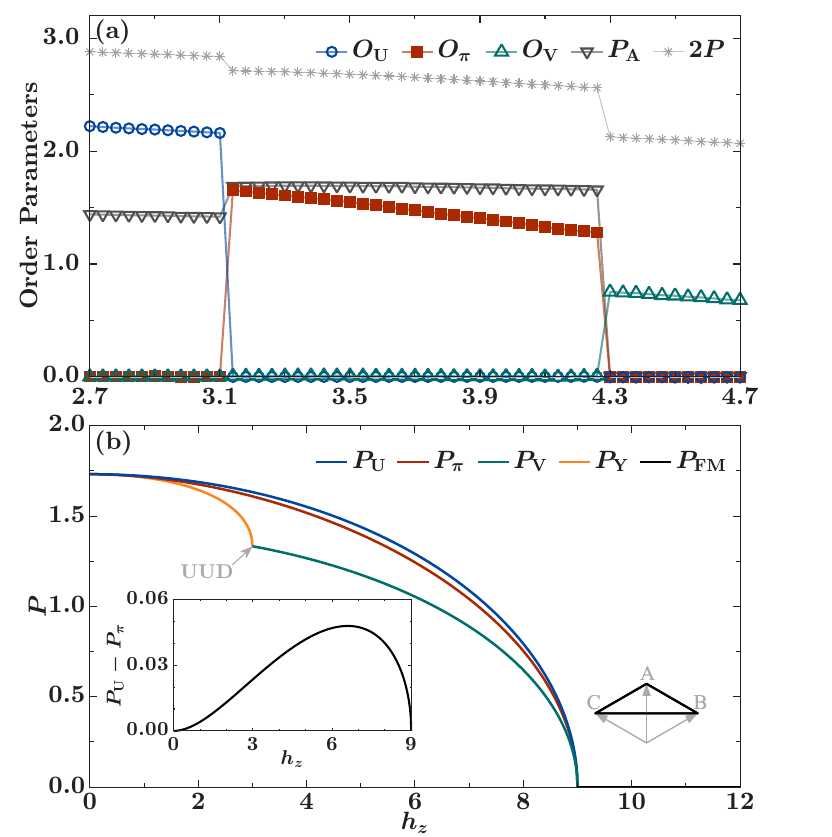}%
    \caption{Order parameters for each phase  at different values of $h_z$. (a) Tensor-network results for $O_U$, $O_\pi$, $O_V$ (for umbrella, $\pi$-coplanar, and V phases) and parity order parameters $P_A$, $P$ at $S=1$, $\eta=0.8$. The $P$ values are doubled to avoid line overlaps. (b) Exact classical results for $P$ of umbrella, $\pi$-coplanar, V, Y, and FM phases at $\eta=1$; the UUD phase is also degenerate at $h_z=3$. The inset shows $P_U-P_\pi$. The lower-right schematic illustrates that $P$ can be represented by the perimeter of the triangle formed by the tips of the $A$, $B$, and $C$ spins.}
    \label{fig:figA2}
\end{figure}

{Furthermore, although our LSW results for $S=1/2$~\cite{supp} are not consistent with our tensor-network calculations, this discrepancy can be attributed to the fact that the $1/S$ quantum corrections in LSW are insufficient to accurately capture the physics of strong quantum fluctuations at $S=1/2$, which also explains why different field-theoretic approaches yield different predictions regarding the location of the $\pi$-coplanar phase.~\mbox{\cite{Yamamoto2014,Starykh2014,Yamamoto2017}}. A more definitive determination for $S=1/2$ requires higher-order corrections combined with more precise numerical validations in the future. Despite the disagreement between our numerical results and mean-field predictions at $S=1/2$, which does not fully settle the question of whether the $\pi$-coplanar phase exists for $S=1/2$, the location where it appears within the LSW level remains consistent with our rule-of-thumb for parity breaking.}

{To quantitatively describe the phase transitions and parity evolution, we compute the order parameters for each phase, including parity order parameters, via tensor-network calculations. Figure~\ref{fig:figA2}(a) presents an example of transitions for $S=1$ and $\eta=0.8$. The order parameters are defined as }
\begin{align}
    O_U &= |\sum_{\langle ij\rangle}(S_i\times S_j)_z|, \\
    O_{\pi} &= |\sum_{\langle ij\rangle}(S_i\times S_j)_x+(S_i\times S_j)_y|, \\
    O_V &= |\sum_{\langle ij\rangle}(S^y_iS^z_j+S^z_iS^y_j+S^z_iS^x_j+S^x_iS^z_j)|
\end{align}
{for the umbrella, $\pi$-coplanar, and V phases, respectively. Here, $O_U$ and $O_\pi$ correspond to the vector chirality along $z$ and along $x$ (or $y$), and $O_V$ measures the nematicity of the V phase.
Parity is captured by both the site-$A$-relative order parameter $P_A=|S_B-S_C|$ and the global average $P=(|S_B-S_C|+|S_C-S_A|+|S_A-S_B|)/3$ over the three sites, with the latter decreasing monotonically with increasing $h_z$. 
After the global parity $P$ is broken, further decreasing $h_z$ strengthens the breaking, for instance through the spontaneous emergence of $P_A$ across the V-to-$\pi$-coplanar transition or V-to-umbrella in Fig.~\ref{fig.phaseDiagram}.
This process bears a conceptual analogy to parity violation in weak interactions: the identical $B$ and $C$ site spins evolve into distinct configurations, reminiscent of the ``$\theta$-$\tau$ puzzle" prior to the work of Lee and Yang, where the $\theta$ and $\tau$ particles were later recognized as different decay modes of the same meson.
We also evaluate $P$ for all degenerate phases at $\eta=1$ in the classical limit [Fig.~\ref{fig:figA2}(b)]. The results also confirm that lowering $h_z$ from the FM state to any phase follows the spontaneous symmetry-breaking scenario. 
}


\newpage

\widetext
\begin{center}
\textbf{\large Supplementary material for \\``Spontaneous parity Breaking in Quantum Antiferromagnets on the Triangular Lattice"}
\end{center}

\setcounter{equation}{0}
\setcounter{figure}{0}
\setcounter{table}{0}

\makeatletter
\renewcommand{\thefigure}{S\arabic{figure}}
\renewcommand{\thetable}{S\arabic{table}}
\renewcommand{\theequation}{S\arabic{equation}}
\renewcommand{\bibnumfmt}[1]{[S#1]}
\renewcommand{\citenumfont}[1]{S#1}
\renewcommand{\bibsection}{} 
\makeatother

\newcolumntype{C}{>{\centering\arraybackslash}X}

\newcommand{\confimg}[2][trim=50pt 40pt 50pt 30pt,clip]{%
  \adjustbox{valign=m}{\includegraphics[width=\linewidth,#1]{#2}}%
}

\newcommand{\vcell}[1]{%
  \adjustbox{valign=m}{#1}%
}

\newcommand{\blocksep}{%
  \specialrule{0.5pt}{1pt}{0.8pt}%
  \specialrule{0.5pt}{1pt}{1pt}%
}

\definecolor{RED}{RGB}{255, 0, 0}
\definecolor{BLUE}{RGB}{0, 0, 255}

\newcommand{\myeq}[1]{ Eq. (\ref{#1}) }
\newcommand{\fig}[1]{Fig.~\ref{#1}}
\newcommand{\pan}[1]{\left\langle #1 \right\rangle}
\newcommand{\pac}[1]{\left\{ #1 \right\}}
\newcommand{\pap}[1]{\left( #1 \right)}
\newcommand{\pas}[1]{\left[#1 \right]}
\newcommand{\paV}[1]{\left\Vert #1 \right\Vert}
\newcommand{\pav}[1]{\left\vert #1 \right\vert}
\newcommand{\pabs}[1]{ \left| #1 \right| }
\newcommand{\vect}[1]{{\bm{#1}}}
\newcommand{\matr}[1]{#1}

\newcommand{\im}{\mathrm{i}}

\newcommand{\id}{\mathds{1}}

\section{(1) Parity properties of different phases}

\begin{figure}[htbp]
\includegraphics[width=1\textwidth]{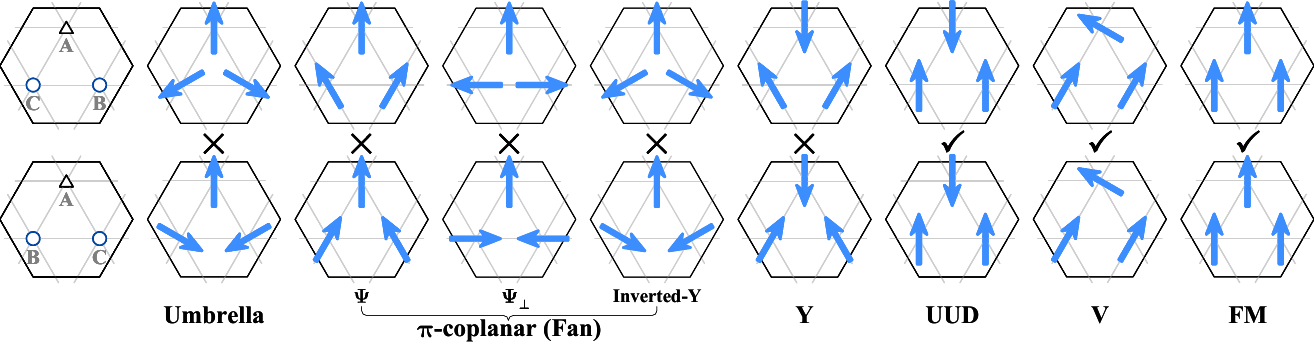}
\caption{
    Schematic illustration of spin configurations [$\vec{S}$=($S_i^x$,$S_i^y$,$S_i^z$), with $i=$ A, B, or C] and their parity symmetry in the monolayer TLXXZ model. Except for the umbrella phase, in which $S_i^z$ is perpendicular to the plane, all other phases are regarded as coplanar phases in the ($S_i^x$, $S_i^z$) plane (with the $S_i^z$ direction chosen as upward). Based on the angle between the spin orientation on site B (or C) and the $S_i^z$ axis, the acute-angle $\Psi$ phase, the right-angle $\Psi_\perp$ phase, and the obtuse-angle inverted-Y phase are collectively referred to as the {$\pi$-coplanar (Fan)} phase {because they are continuously connected without intervening phase boundaries~\cite{SYamamoto2019}}. The upper and lower panels show the spin configurations before and after the parity transformation centered at site A, respectively. Among these phases, the umbrella, {$\pi$-coplanar}, and Y phases break parity (indicated by $\times$), while the UUD, V, and FM phases preserve parity (indicated by $\checkmark$).
}
\label{fig.paritySchem}
\end{figure}

The ground states of the monolayer antiferromagnetic (AF) triangular-lattice XXZ (TLXXZ) model have been extensively studied in the literature. The same phase often appears under different names depending on the perspective adopted. For instance, the {$\pi$-coplanar} phase~\cite{SYamamoto2014} is also referred to as the $\Psi$ phase~\cite{SStarykh2014} or the {Fan} phase~\cite{SMaryasin2013}. Notably, the {$\pi$-coplanar} phase and the Inverted-Y phase are adiabatically connected in the easy-axis limit~\cite{SYamamoto2019}. From the perspective of geometric configurations, their only difference lies in whether the deviations of the two mirror-symmetric spin orientations from the external field are acute or obtuse angles. When this deviation is exactly a right angle, it can also be denoted as $\Psi_\perp$ phase (the special nature of this case will be addressed later in this section). Therefore, in this work, we collectively refer to these three cases as the {$\pi$-coplanar} phase. Meanwhile, the V phase is alternatively known as the 0-coplanar phase~\cite{SYamamoto2014}. The spin configurations of the phases mentioned above, as well as those of the other ground states in the monolayer TLXXZ model, can be displayed within a unit cell composed of A, B, and C sites (Fig.~\ref{fig.paritySchem}).

More specifically, each ground state and the corresponding phase transitions can be characterized by the magnetization $M_z=\langle S_A^z+S_B^z+S_C^z\rangle/3$ or other order parameters. The UUD and FM phases exhibit a 1/3 magnetization plateau and saturation, respectively, whereas the remaining phases display a gradual increase of 
$M_z$ with increasing field, which can be further distinguished by defining appropriate correlations or chiralities~\cite{SXu2012}. Nevertheless, given that all these phases are ordered, the angle $\theta_{ij}$ among the spins on $i,j=$ A, B, and C sublattice sites within one unit cell are in fact sufficient to fully distinguish all the aforementioned phases. These angles can be directly extracted from the computed $\langle S_i^{\alpha}\rangle$ ($\alpha=x,y,z$) expectation values. In the umbrella phase, the angles satisfy $\theta_{AB}=\theta_{BC}=\theta_{CA}>0$; in the {$\pi$-coplanar} phase, $\theta_{BC}=2\theta_{AC}\neq 0$; in the Y phase, $\theta_{BC}=2(\pi-\theta_{AC})\neq 0$; and in the V phase, $\theta_{BC}=0,\theta_{AC}\neq 0$.

Once the spin configurations are obtained, they can be analyzed from a symmetry perspective. Specifically, the spins in the umbrella and {$\pi$-coplanar} phases are arranged in different directions around the magnetic field direction, exhibiting strong rotational or mirror symmetry. In contrast, in the V phase, two spins are aligned and tilted away from the magnetic field, so the configuration is mirror-asymmetric with respect to the field direction. However, the symmetry classification changes from the perspective of parity. Since parity corresponds to spatial inversion and the spin is a pseudovector, a parity transformation leaves the spin direction unchanged and only permutes the lattice sites on which the spins reside. In the unit cell of the monolayer TLXXZ system, if site A is chosen as the inversion center, the parity transformation simply exchanges the sites B and C, as shown in the Fig.~\ref{fig.paritySchem}. Under this operation, the spin configurations of the umbrella and {$\pi$-coplanar} phases are transformed into different configurations, indicating spontaneous parity nonconservation. In contrast, the V phase, although geometrically less symmetric at first sight, preserves parity symmetry. Similarly, the Y phase breaks parity upon transformation, whereas the UUD and FM phases preserve parity conservation.

In the main text, we point out that all the above phases in the monolayer TLXXZ model, as the external field $h_z$ increases, must satisfy a rule-of-thumb parity principle. Specifically, under strong magnetic fields, the ground state tends to preserve parity symmetry, whereas under weak fields, parity symmetry tends to be spontaneously broken. This rule is verified by tensor-network data and can be qualitatively understood through the competition between $h_z$ and the $S_i^zS_j^z$ term in the model. In the End Matter, we construct effective Hamiltonians for all phases via mean-field theory and further validate this rule using linear spin-wave theory. 

If one gradually reduces $S$ from this limiting case, the entire {$\pi$-coplanar} phase shifts into the $\eta<1$ regime, in full agreement with both our numerical and mean-field results. Moreover, our tensor-network numerical calculations further reveal both $\Psi_\perp$ and obtuse {$\pi$-coplanar} (Inverted-Y) configurations inside the entire {$\pi$-coplanar} region. Specifically, the $\Psi_\perp$ configuration appears at $S=1$, $\eta=0.90$, $h_z/S=2.8$ 
 and at $S=5/2$, $\eta=0.96$, $h_z/S=3.0$, whereas the obtuse {$\pi$-coplanar} configuration appears at $S=1$, $\eta=0.94$, $h_z/S=2.1$ with $\theta_{AB}=98^\circ$. These results not only substantiate the validity of the {$\pi$-coplanar} phase region obtained in this work, but also verify that the three branches with different angles remain adiabatically connected in the easy-plane limit.
 
Beyond serving as a basis for validating the parity guiding principle, the {$\pi$-coplanar} phase region studied in this work is of interest in its own right. For example, in experiments on TLXXZ-type materials, the physical properties and excitation spectrum of the {$\pi$-coplanar} phase remain largely unexplored, possibly because inducing the {$\pi$-coplanar} phase has generally been thought to require strong magnetic fields. This work demonstrates that some large-$S$ materials may realize the add{$\pi$-coplanar} phase under moderate external fields. In addition, although $\eta$ is a tunable parameter in calculations, determining the anisotropy parameter in real triangular-lattice compounds can be complicated. This work shows that, beyond the existence of the UUD phase, the presence of the {$\pi$-coplanar} phase can also provide a useful reference for constraining the approximate range of $\eta$ in real materials. Moreover, the significance of the {$\pi$-coplanar} phase is complementary to that of the V phase. In the bosonic language, the {$\pi$-coplanar} phase exhibits properties distinct from those of the V phase, which is a typical supersolid. The emergence of the {$\pi$-coplanar} phase provides an alternative picture of the superfluid-to-supersolid transition, distinct from the direct umbrella-to-V transition. Therefore, the outcome of the competition between the {$\pi$-coplanar} and V phases, as well as the connection between this competition and their superfluid properties, remains an open question worthy of further investigation.

\section{(2) Details of the tensor-network calculation}

Our tensor-network calculations mainly employ the projected entangled simplex
state (PESS) algorithm~\cite{SXie2014} to obtain the ground states, and utilize the modified corner-transfer-matrix renormalization
group (CTMRG) method mentioned in the main text to compute physical quantities such as energies, spin expectation values, and magnetization along the external field. A detailed description of these two algorithms is provided below.

\subsection{A) The PESS ansatz}

We employ the PESS ansatz to obtain the ground states of the TLXXZ model. Broadly speaking, this method discretizes the imaginary-time evolution operator into local evolution operators, which are then used to iteratively update all tensors, including simplex tensors, bond vectors, and local tensors with physical indices, until convergence. Specifically, for sufficiently large $\beta$, the imaginary-time evolution operator $U_\beta=e^{-\beta H}$ projects a suitable initial state onto the ground state, where $H$ is the Hamiltonian [e.g., Eq. (1) in the main text for the monolayer system] and $\beta$ denotes the imaginary time. Using the Trotter decomposition, $U_\beta$ can be discretized into a sequence of imaginary-time evolution steps $U_\tau =e^{-\tau H}$ with a small time step $\tau$, which is set to $\tau=0.01$ in this work. In addition to the decomposition in imaginary time, the Hamiltonian $H$ can be further decomposed according to the upward triangles of the lattice:

\begin{figure}[htbp]
\includegraphics[width=0.7\textwidth]{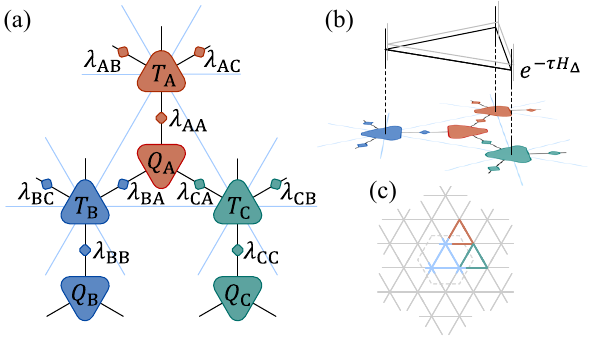}
\caption{
    Schematic illustration of the PESS algorithm for the TLXXZ model. (a) Wave-function ansatz: local tensors $T_i$ (upward triangles), where $i=A,B,C$ labels the sites within a unit cell, are placed on the sites of the triangular lattice (blue lines) and are connected to simplex tensors $Q_i$ (downward triangles) through bond vectors $\lambda_{ij}$ (small squares). (b) Local imaginary-time evolution: the local evolution operator $e^{-\tau H_\Delta}$ of monolayer system is represented as a matrix-product operator (black upward triangle) and applied to the wave-function tensors. (c) The local evolution operator is applied successively to three local tensor structures composed of $A$, $B$, and $C$, which are represented by upward triangles in three different colors.
}
\label{fig.SimpleUpdate}
\end{figure}

\begin{align}
\label{eq.H.supp}
H &= \sum_{\Delta} H_{\Delta}, \\
\label{eq.H.Delta}
H_{\Delta} &= J_{ab} \sum_{\braket{i,j}} \pap{S_i^x S_j^x + S_i^y S_j^y + \eta S_i^z S_j^z} - \frac{h_z}{3} \sum_i S_i^z,
\end{align}
where the index $\Delta$ labels each upward triangle and the indices $i$ and $j$ in $H_\Delta$ denote the three vertices of the triangle $\Delta$. Note that the summation over the bonds of all $\Delta$ in Eq.~(\ref{eq.H.supp}) naturally includes all nearest-neighbor interactions. However, if the field term $h_z S_i^z$ were summed directly over the vertices of all $\Delta$, it would be counted three times in Eq.~(\ref{eq.H.supp}). Therefore, the field strength $h_z$ is reduced by a factor of three in Eq.~(\ref{eq.H.Delta}). With these decompositions in imaginary time and in the Hamiltonian, $U_\beta$ is discretized into local evolution matrices $e^{-\tau H_\Delta}$, which can be constructed as sixth-order matrix-product operators (MPOs) through tensor permutation and reshaping.

Based on the PESS ansatz, the wave function of the TLXXZ model is constructed as a tensor network shown in Fig.~\ref{fig.SimpleUpdate}(a):
\begin{equation}
\label{eq.PESS.ansatz}
\ket{\Psi} = \text{Tr} \pap{ \cdots T^\alpha_{i} \lambda_{ij} Q_{j} \cdots} \ket{\cdots \alpha \cdots},
\end{equation}
where $T^\alpha_i$ is the local tensor on site $i$ ($i=A,B,C$) in the unit cell, with $\alpha$ denoting the physical spin degree of freedom on that site; $Q_j$ is the simplex tensor associated with site $j$; and the matrix $\lambda_{ij}$ is the bond vector surrounding $T^\alpha_i$ and connecting it to $Q_j$. The trace Tr denotes the contraction over all virtual indices, and the bond dimension used in this work is $D_b=6$. The local evolution can then be implemented by applying $e^{-\tau H_\Delta}$ successively to the upward triangles centered at different $Q_i$ tensors. After each application, a Tucker decomposition is performed to update $Q_i$, the connected matrices $\lambda_{ji}$, and all local tensors $T_i$. For example, $e^{-\tau H_\Delta}$ is first applied to the upward triangle centered at $Q_{\rm A}$, updating $Q_{\rm A}$, $\lambda_{\rm AA}$, $\lambda_{\rm BA}$, $\lambda_{\rm CA}$, and all local tensors, as shown in Fig.~\ref{fig.SimpleUpdate}(b). The $T$ tensors and $\lambda$ matrices are then cyclically permuted to form new upward triangles centered at $Q_{\rm B}$ and $Q_{\rm C}$, which are evolved in turn. Spatially, this procedure corresponds to evolving three neighboring upward triangles one by one, as illustrated in Fig.~\ref{fig.SimpleUpdate}(c). After sufficient iterations, the wave function is projected onto the ground states.

For the bilayer system, the Hamiltonian can be written as
\begin{align}
H_\text{bilayer} = H^{(1)} + H^{(2)} + H',
\end{align}
where $H^{(1)}$ and $H^{(2)}$ denote the intralayer interactions in the first and second layers, respectively, as defined in Eq.~(1) in the main text, and $H'$ denotes the interlayer interaction defined in Eq.~(2) in the main text. Similarly, $H'$ can be decomposed according to the upward triangles:
\begin{align}
\label{eq.H'.supp}
H' &= \sum_{\Delta} H_\Delta', \\
H_\Delta' &= \frac{J_c}{3} \sum_i \vec{S}_i^{(1)} \cdot \vec{S}_{i}^{(2)} + J_d \sum_{\braket{i,j}} \pap{ \vec{S}_i^{(1)} \cdot \vec{S}_{j}^{(2)} + \vec{S}_{i}^{(2)} \cdot \vec{S}_j^{(1)} },
\end{align}
where the superscripts $(1)$ and $(2)$ label the first and second layers, respectively, and the sites $i$ and $j$ run over the three vertices of a given $\Delta$. To avoid overcounting the $J_c$ term, $J_c$ is reduced by a factor of three in $H_\Delta'$. Combining Eqs.~(\ref{eq.H.supp}) and (\ref{eq.H'.supp}), $H_\text{bilayer}$ can also be decomposed according to the upward triangles $\Delta$. The resulting bilayer local evolution operator can be brought into the same structure as the monolayer MPO by permuting and reshaping tensor indices, with the two layer degrees of freedom on the same site reshaped into a single physical index, as illustrated in Fig.~\ref{fig.SimpleUpdate}(b). For example, the dimension of the physical indices of bilayer system with $S=1/2$ is equivalent to that of monolayer system with $S=3/2$. In short, the PESS scheme described above for the monolayer system can be naturally extended to the bilayer TLXXZ model.

\subsection{B) The modified CTMRG contraction scheme}

\begin{figure}[htbp]
\includegraphics[width=0.6\textwidth]{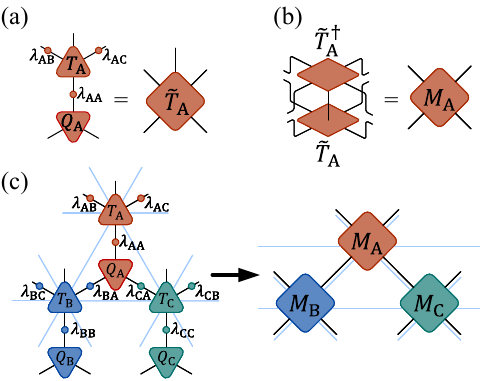}
\caption{
    Transformation of the hexagonal wave-function ansatz into a square-network structure. (a) The local tensor, simplex tensor, and bond vectors associated with each site $i$ are contracted into a rank-5 tensor $\tilde{T}_i$ (take site A for example). (b) $\tilde{T}_i$ is contracted with $\tilde{T}_i^\dagger$, the physical index is traced out, and the resulting tensor is reshaped into a rank-4 tensor $M_i$. (c) The tensors $M_i$ form a square-network structure. 
}
\label{fig.Triangle.Square}
\end{figure}

The local tensors obtained by the last step encode only the microscopic local properties of the system. To evaluate macroscopic observables, such as the energy and the expectation values of spin operators, one has to incorporate the environment of the infinite lattice using tensor-renormalization methods. And CTMRG~\cite{SNishino1996,SCorboz2010} provides such an efficient renormalization strategy. For generality and conciseness, we perform CTMRG in a standard square-network pattern. To this end, the tensors in the wave-function ansatz are reorganized. Specifically, for each site $i$, the local tensor $T_i$, the environment matrices $\lambda_{i,j}$, and the simplex tensor $Q_i$ are contracted over their shared virtual indices to form a rank-5 tensor $\tilde{T}_i$, as shown in Fig.~\ref{fig.Triangle.Square}(a). Then, $\tilde{T}_i$ is contracted with its Hermitian conjugate $\tilde{T}_i^\dagger$ over the physical index, yielding a rank-4 local tensor $M_i$, as shown in Fig.~\ref{fig.Triangle.Square}(b). In this way, the PESS ansatz is reorganized into a square lattice tensor network, as illustrated in Fig.~\ref{fig.Triangle.Square}(c).
This transformation amounts only to a graphical rearrangement of the triangular lattice (blue lines) and does not affect the physical properties of the system.

Instead of assigning the same set of environment tensors to all local tensors, we assign an independent set of environments to each $M_i$, namely edge tensors $L_i^\mu$ with $\mu=l,d,r,u$, and corner tensors $L_i^{\mu\nu}$ with $\mu\nu=lu,ld,rd,ru$. These tensors $L_i^\mu$ and $L_i^{\mu\nu}$ can be directly contracted with $M_i$ to form an approximation to the infinite system viewed from the position of a single $M_i$. This approximation can also be decomposed by grouping several $M_i$ tensors together and using the corresponding $L_i^\mu$ and $L_i^{\mu\nu}$ tensors as their surrounding environment. In this way, all three connectivity patterns of the TLXXZ model can be constructed, as illustrated in Fig.~2 in the main text. For unit cells with more sites, the same strategy can be used to construct larger connectivity patterns.

\begin{figure}[htbp]
\includegraphics[width=1\textwidth]{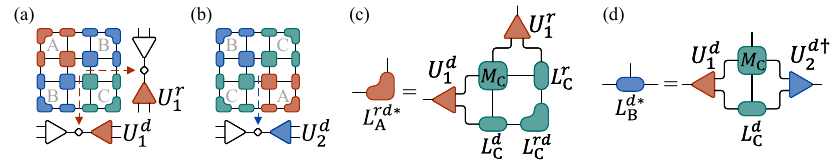}
\caption{
    Example of updating corner and edge environment tensors. (a) The right and bottom cuts of block (1) in Fig.~2(a)(the main text) generate $U_1^r$ and $U_1^d$, respectively. (b) The bottom cut of block (2) corresponding to Fig.~2(b) generates $U_2^d$. (c) Update of the lower-right corner tensor $L_{\rm A}^{rd*}$ of $M_{\rm A}$. (d) Update of the bottom edge tensor $L_{\rm B}^{d*}$ of $M_{\rm B}$.
}
\label{fig.CTMRG.gateContr}
\end{figure}

The CTMRG procedure based on these connectivity patterns has good parallelizability, because we do not need to evolve separately along the four directions (left, right, up, and down); instead, all Blocks in Fig.~2 (in the main text) can be constructed and the four corresponding $U$ tensors can be obtained simultaneously, and there is no strict sequential constraint requiring certain isometries to be constructed before others. Therefore, the construction of isometries and the update of individual environment tensors can be performed in parallel. For instance, the SVDs associated with the bottom and right cuts of block (1), as well as the bottom cut of block (2), can be carried out simultaneously, yielding $U_1^d$, $U_1^r$, and $U_2^d$, respectively, as shown in Figs.~\ref{fig.CTMRG.gateContr}(a) and \ref{fig.CTMRG.gateContr}(b). Notably, the independence and parallelizability of the environment tensors in block construction do not imply that the tensors $L_i^\mu$ and $L_i^{\mu\nu}$ are unrelated. On the contrary, environment tensors associated with different sites are transformed into one another after the update. For example, after one coarse-graining step, $L_{\rm C}^{rd}$ corresponds to the updated environment tensor $L_{\rm A}^{rd*}$ of $M_{\rm A}$, as shown in Fig.~\ref{fig.CTMRG.gateContr}(c), while the coarse graining of $L_{\rm C}^d$ gives $L_{\rm B}^{d*}$, as shown in Fig.~\ref{fig.CTMRG.gateContr}(d). The basic idea of these coarse-graining steps is to truncate the isometries $U$ obtained from the cuts, insert $U^\dagger U\approx \id$ back into the cuts, where $\id$ is the identity matrix, and contract them with the target tensors, thereby keeping the virtual dimension of the tensor network fixed at $\chi$.

The updated tensors $L_i^{\mu*}$ and $L_i^{\mu\nu*}$ serve as the environment tensors for the next iteration and are used to reconstruct the three blocks. By repeatedly applying the procedure shown in Fig.~2 in the main text, the environment tensors eventually converge. With the converged environment tensors, one can evaluate the expectation value of the spin operator on each site,
\begin{align}
\braket{\vec{S}_i} = \frac{\bra{\Psi} \vec{S}_i \ket{\Psi}}{\braket{\Psi|\Psi}},
\end{align}
as well as {correlations (e.g., energy, chirality, and nematicity) with the form},
\begin{align}
{C_{ij}} = \frac{\bra{\Psi} {S^\alpha_iS^\beta_j} \ket{\Psi}}{\braket{\Psi|\Psi}}.
\end{align}

\begin{figure}[htbp]
\includegraphics[width=0.7\textwidth]{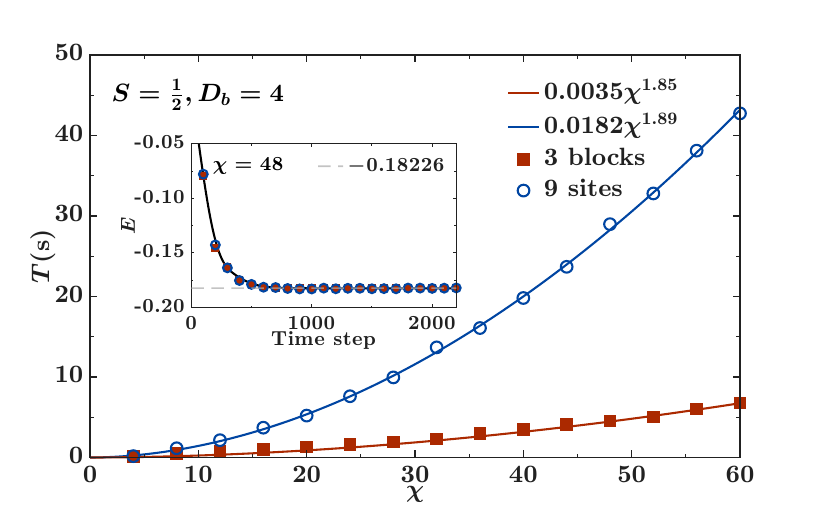}
\caption{
    Comparison of the computational time and energy results obtained from two CTMRG strategies. Take the monolayer TLXXZ model at $S=1/2, \eta=1$ as an example with $D_b=4$. The red squares and blue circles represent the modified CTMRG contraction strategy used in this work and a conventional CTMRG contraction strategy~\cite{SLi2022}, respectively. The red and blue lines show fits to the computational time of the two strategies for the same ground state. The inset shows the evolution of the energies calculated by the two strategies during the imaginary-time evolution from the same random initial state to the ground state. The black line is a guide to the eye, and the gray dashed line marks the reference ground state energy $E_0=-0.18226$ reported in Ref.~\cite{SLi2022}.
}
\label{fig.E.bench.CTMRG.old.nr}
\end{figure}

For comparison, we also implemented a representative CTMRG strategy based on the connectivity pattern used in Ref.~\cite{SLi2022}. We benchmark our numerical results and computational time with those from Ref.~\cite{SLi2022}. Using a virtual bond dimension $D_b = 4$ for the local tensors (with CTMRG truncation dimension
$\chi$ = 48), we obtain a ground-state energy of $E_0$ =-0.1824 for the triangular-lattice AF Heisenberg model at zero field, in agreement with results obtained from Ref.~\cite{SLi2022}. The average computational time for one CTMRG step of the two strategies for different values of $\chi$ are shown in Fig.~\ref{fig.E.bench.CTMRG.old.nr}. The results show that the CTMRG strategy developed in this work has an advantage in computational speed. For example, at $\chi=60$, the computational speed of the modified version is 6.3 times that of the conventional one. It is worth noting that we have not employed other parallelization techniques such as GPU acceleration in the current work. It is worth noting that we have not employed other parallelization techniques such as GPU acceleration in the current work. In principle, apart from the transfer of intermediate data, the calculations for the three columns in Fig.~2 of the main text are independent of each other. If further parallelization were implemented at the hardware level or within the CPU programming, the acceleration provided by the modified method would be even more pronounced. Meanwhile, the two CTMRG strategies yield consistent energy evolution during the imaginary-time evolution, which confirms the validity of the CTMRG strategy developed in this work.

\section{(3) Details of the numerical results}
\subsection{A) Comparing energies to determine phase boundaries}

Based on the PESS method and CTMRG scheme described above, we obtain different spin configurations and compute their energies, thereby determining the ground-state phase diagram. The basic procedure for constructing the phase diagram is as follows. First, within each phase region, we select a point and start from a random initial wave function, then perform imaginary-time evolution to obtain a stable wave function corresponding to the relevant spin configuration. Next, starting from the previous stable wave function, we gradually tune the parameters $h_z$ and $\eta$ and continue the imaginary-time evolution, allowing the wave function to relax into the state corresponding to the new parameters, until the phase boundary is reached or crossed. During this parameter sweep, two situations may occur. In the first case, a given spin configuration becomes unstable under imaginary-time evolution and evolves into another configuration, indicating that the original configuration is no longer the ground state. In the second case, multiple configurations remain stable at the same point even after sufficiently long imaginary-time evolution. In this case, the ground state at that point is identified as the configuration with the lowest energy.

\begin{figure}[htbp]
\includegraphics[width=1\textwidth]{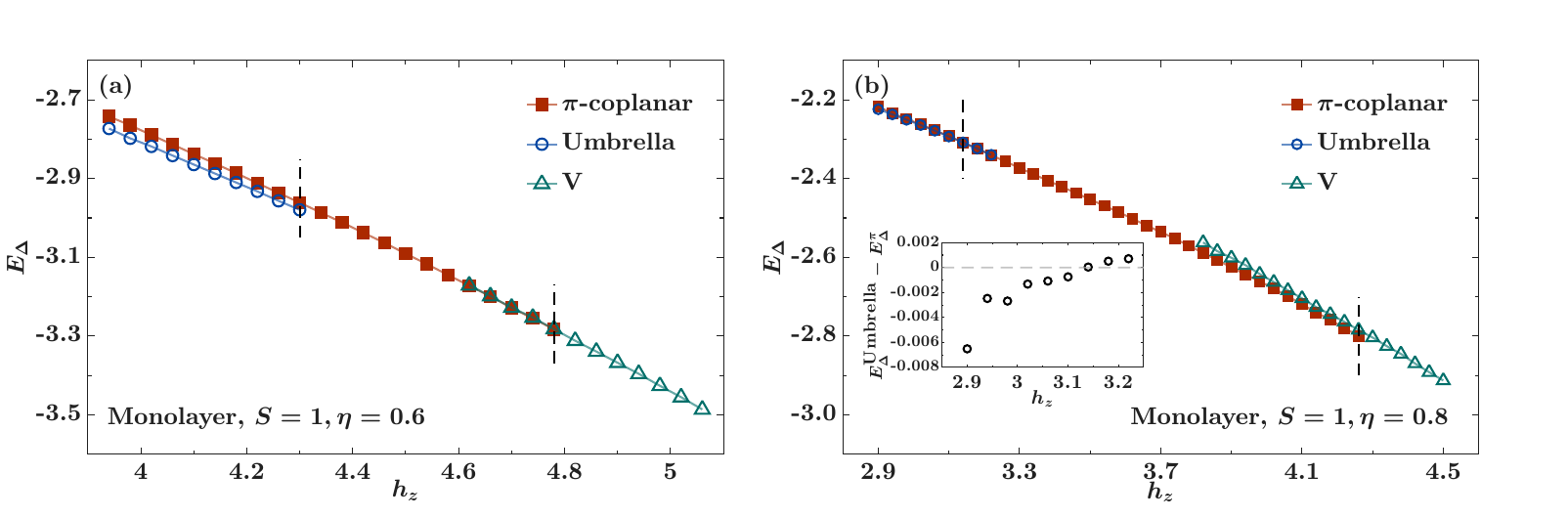}
\caption{
    Energy comparison in the monolayer system. The blue circles, red squares, and green triangles denote the energies of the umbrella, {$\pi$-coplanar}, and V states, respectively. The vertical dashed lines mark the positions where the ground state phase changes. Panels (a) and (b) show the results for $\eta=0.6$ and $\eta=0.8$, respectively, with $D_b=6, \chi=72$. The inset in panel (b) shows the energy difference between the umbrella ($E_\Delta^\text{Umbrella}$) and {$\pi$-coplanar} ($E_\Delta^\text{{$\pi$}}$) configurations.
}
\label{fig.E.crossover}
\end{figure}

As an example, we examine the existence of the {$\pi$-coplanar} phase at $S=1$ for $\eta=0.6$ and $\eta=0.8$. We first obtain well-converged wave functions for the {$\pi$-coplanar}, umbrella, and V configurations deep inside their respective phase regions. We then gradually vary the external field $h_z$, obtain the stable states of each spin configuration at a series of $h_z$ values, and compute their energies. For $\eta=0.6$, the umbrella and {$\pi$-coplanar} configurations both remain stable for $h_z\leq 4.30$, while the umbrella configuration has a lower energy, indicating that the umbrella state is the ground state, as illustrated in Fig.~\ref{fig.E.crossover}(a). Once $h_z$ becomes slightly larger than 4.30, the umbrella configuration cannot remain stable after sufficiently long imaginary-time evolution, and the ground state changes to the {$\pi$-coplanar} phase. As $h_z$ is further increased, the {$\pi$-coplanar} and V configurations both remain stable, and the {$\pi$-coplanar} configuration has a lower energy for $h_z\leq 4.78$. Once $h_z$ becomes slightly larger than 4.78, the {$\pi$-coplanar} configuration also becomes unstable after sufficiently long imaginary-time evolution, and the ground state changes to the V phase. In fact, in the range $4.34\lesssim h_z\lesssim 4.5$, neither the umbrella nor the V configuration remains stable after sufficiently long imaginary-time evolution; in this regime, the stability under imaginary-time evolution alone is sufficient to identify the existence of the {$\pi$-coplanar} phase as a ground state.

In addition to the unstability of a given configuration during imaginary-time evolution, phase transitions can also be identified through energy crossover. For example, at $\eta=0.8$ and around $h_z\sim 3.14$, the energy of the umbrella phase becomes higher than that of the {$\pi$-coplanar} phase, indicating a ground-state transition from the umbrella phase to the {$\pi$-coplanar} phase, as illustrated in Fig.~\ref{fig.E.crossover}(b). Similar to the previous case, the {$\pi$-coplanar} phase remains stable up to $h_z = 4.26$; although it coexists with the V configuration, its energy is lower than that of the V state. When $h_z$ is slightly increased from 4.26, the {$\pi$-coplanar} configuration becomes unstable under imaginary-time evolution, indicating that the ground state changes to the V phase. These results quantitatively demonstrate that the {$\pi$-coplanar} phase occupies a region between the umbrella and V phase regions.

\subsection{B) Numerical results for the bilayer TLXXZ model}

Interlayer interactions introduce additional competing exchange paths, leading to richer frustrated spin configurations and phase behavior. A typical example is the field-induced magnetization process, in which dimers are successively broken, resulting in a series of magnetization plateaus at increasing magnetizations. This magnetization process involves rich changes in spin configurations and phase-diagram structures. The main text presents representative magnetization processes and spin configurations (Fig. 4) and discusses the parity symmetry of these configurations. This subsection further describes and discusses these spin configurations in detail with representative numerical examples.

\begin{table}[htbp]
    \centering
    \caption{Representative spin configurations of the bilayer TLXXZ model at different $h_z$ (Phase abbreviations follow Fig. 4 in the main text). The three-dimensional coordinates of each arrow represent the expectation values of the corresponding spin-operator components, and the arrow length represents the magnitude of the spin. The configurations correspond to the schematic insets in Fig.~4. All spins are coplanar, and the arrows are drawn in the $x$-$z$ plane.
}
    \begin{tabular}{ccccc}
        \toprule
        $h_z$ (State) 
        & 0.7 (Dimer) 
        & 1.1 (BEC1) 
        & 1.2 (NS1) 
        & 2.1 (SS1) \\
        \midrule
        Configuration &
        \includegraphics[width=3cm, valign=m, trim=50pt 40pt 50pt 30pt,clip]{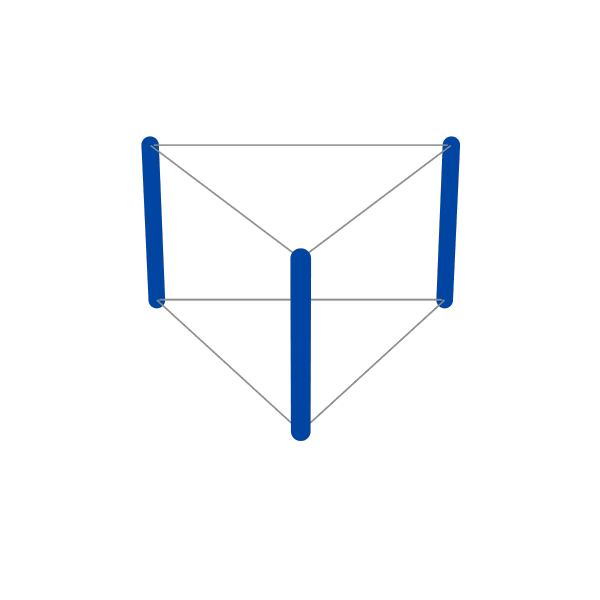} &
        \includegraphics[width=3cm, valign=m, trim=50pt 40pt 50pt 30pt,clip]{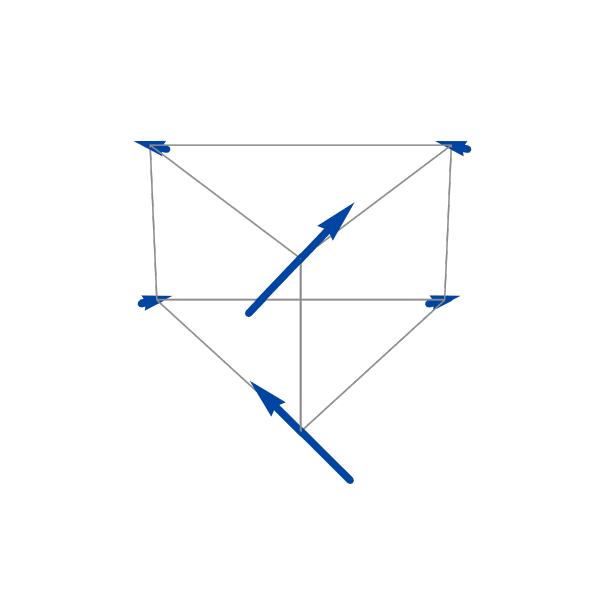} &
        \includegraphics[width=3cm, valign=m, trim=50pt 40pt 50pt 30pt,clip]{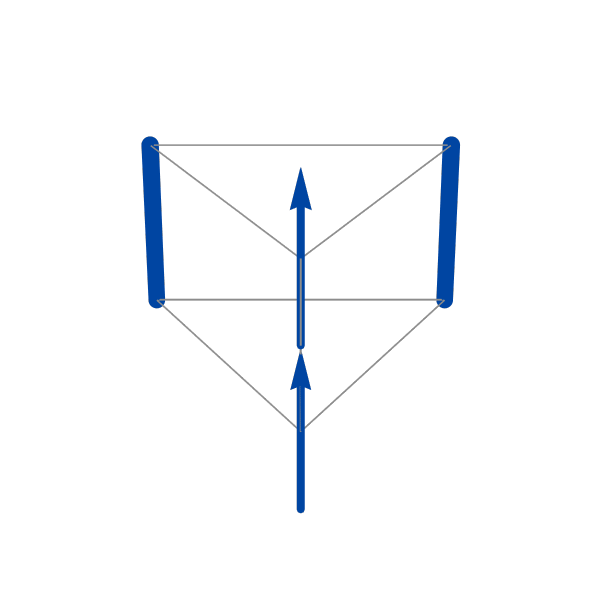} &
        \includegraphics[width=3cm, valign=m, trim=50pt 40pt 50pt 30pt,clip]{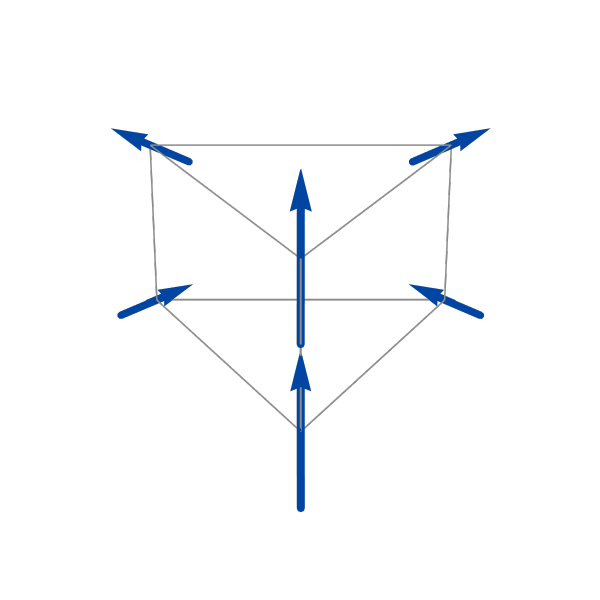}
        \\
        \blocksep
        $h_z$ (State) 
        & 2.2 (SS2) 
        & 2.3 (NS2) 
        & 3.3 (BEC2) 
        & 4.0 (FM) \\
        \midrule
        Configuration &
        \includegraphics[width=3cm, valign=m, trim=50pt 40pt 50pt 30pt,clip]{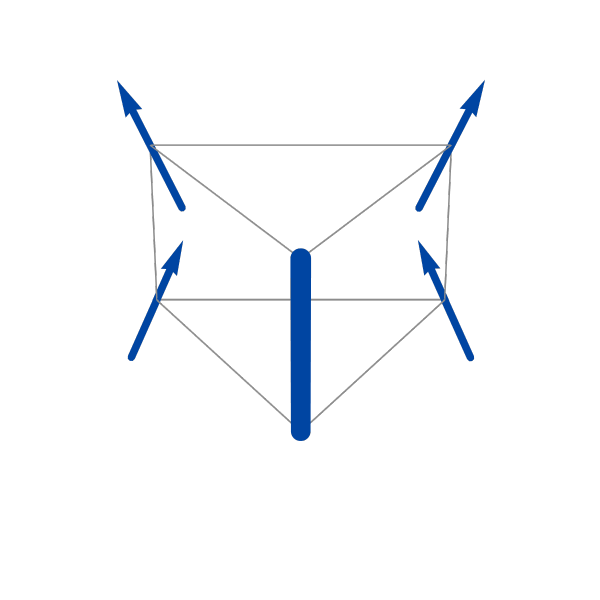} &
        \includegraphics[width=3cm, valign=m, trim=50pt 40pt 50pt 30pt,clip]{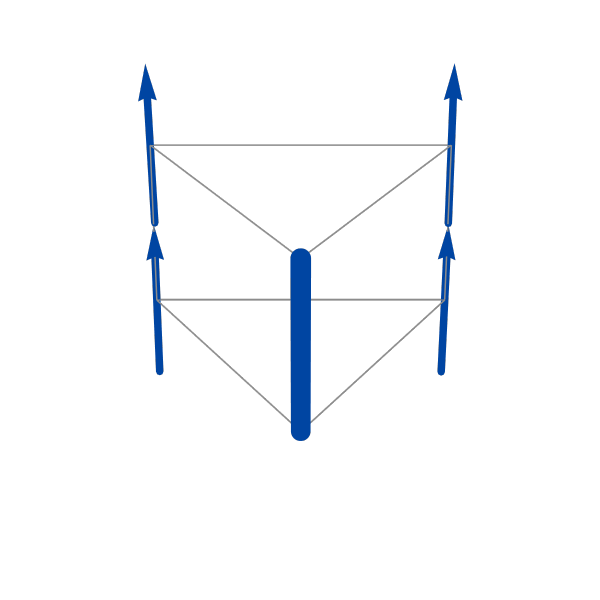} &
        \includegraphics[width=3cm, valign=m, trim=50pt 40pt 50pt 30pt,clip]{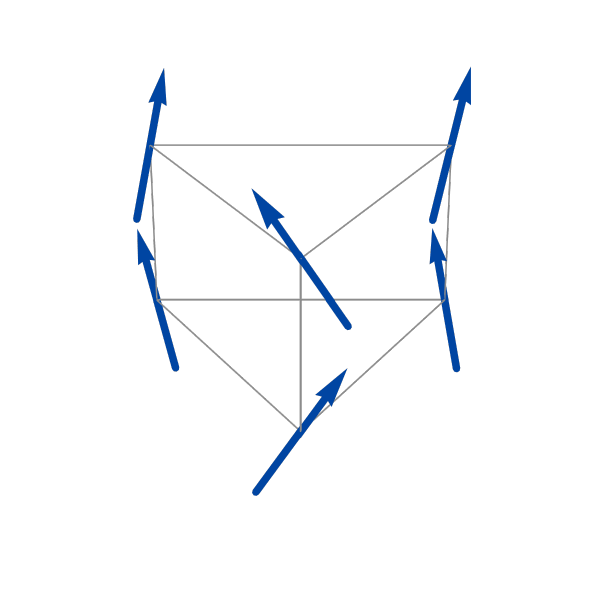} &
        \includegraphics[width=3cm, valign=m, trim=50pt 40pt 50pt 30pt,clip]{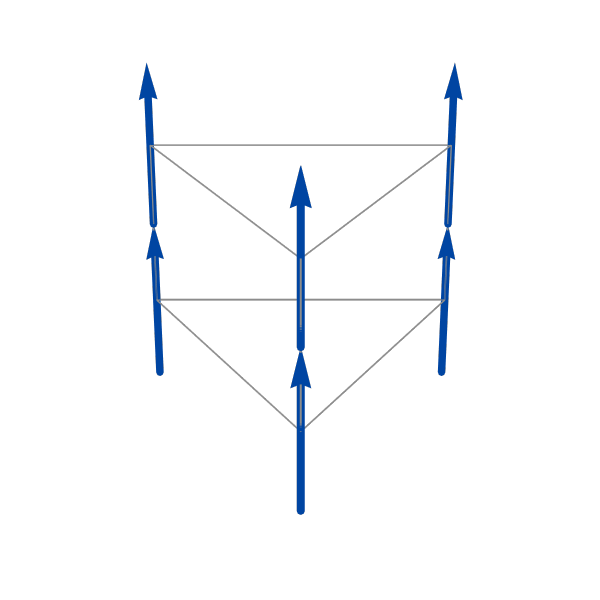}
        \\
        \bottomrule
\end{tabular}

\label{tab.bilayerConf}
\end{table}

As in the monolayer case, we determine the spin configurations by calculating the expectation values of the spin operators, and identify the ground state at a given $h_z$ through energy comparison, with representative examples shown in Table~\ref{tab.bilayerConf}. At weak magnetic field $h_z$, spins at different layers form singlet dimers. The expectation values of the spin components of these dimers are nearly zero, as shown by the state Dimer in Table~\ref{tab.bilayerConf}, giving rise to the first magnetization plateau. As $h_z$ increases, one of the three dimers is strongly broken into a partially polarized spin pair, while the remaining two pairs slightly deviate from the dimer state and acquire small spin components (BEC1). Upon further increasing $h_z$, the polarized pair becomes fully aligned with the field, whereas the other two pairs return to the dimer state, resulting in the $M_s/3$ plateau with $M_s=1/2$, as illustrated by the state NS1. Beyond the $M_s/3$ plateau, the fully polarized pair remains unchanged, while the other two pairs again depart from the dimer state, forming a Fan-type configuration in each layer (SS1). Then, one spin pair returns to the dimer state, while the other two pairs completely lose their dimer character (SS2). These two polarized pairs subsequently become saturated, forming the $2M_s/3$ magnetization plateau (NS2). The remaining dimer is eventually broken at strong $h_z$, leading to a V-type configuration in each layer (BEC2). Finally, the system reaches full saturation (FM).

During this process, the total number of well-defined singlet (S) and triplet (T) pairs remains fixed in the plateau states, whereas such well-defined pairs are absent in the BEC states. Notably, the triplet pair survives in the SS1 phase after the NS phase, and the positions of the singlet and triplet pairs can be different between the SS1 and SS2 phases. Meanwhile, rather than simply remaining aligned, the intralayer spins tend to form configurations similar to those in the monolayer system, such as V-type configurations in the BEC phase and Fan-type configurations in the SS phases. Therefore, the bilayer system naturally exhibits parity symmetry analogous to that of the monolayer system, such as Parity a. In the plateau and BEC phases, two spin pairs have identical configurations, so Parity a is preserved. However, this is not the case for the SS phases. Within each layer of the SS phases, two spin pairs are symmetric with respect to the magnetic-field direction; the same symmetry also holds for the two spins within each of these pairs. This means that the SS phases are invariant under a spatial inversion that simultaneously exchanges the intralayer and interlayer spins, namely Parity b. By contrast, Parity b is broken in the BEC phase. 

Compared with the monolayer case, interlayer interactions introduce an additional parity symmetry behavior that is absent in the monolayer system. These observations highlight parity as a useful perspective for analyzing frustrated systems. Systems with multiple competing interactions may exhibit richer parity-symmetry structures. Identifying the most physically relevant parity operation among several candidates may then become an important question in the analysis of frustrated systems.

\subsection{C) Effects of $\eta$ and coupling on magnetization plateaus}

A conventional order parameter to characterize magnetization-plateau phases, such as the UUD phase, is the magnetization,
\begin{align}
M_z=\frac{1}{n} \sum_{i}^{n} \braket{S_{i}^z},
\end{align}
where $n$ is the number of spins in the unit cell. In the monolayer TLXXZ model with $n=3$, the isotropic case with $S=1$ and $\eta=1$ exhibits the typical $1/3$ magnetization plateau of the UUD phase, as illustrated in Fig.~\ref{fig.Mz}(a). As $\eta$ decreases, the width of the magnetization plateau induced by the longitudinal field gradually narrows and eventually disappear. For example, for $S=1$ and $\eta=0.8$, the UUD phase is absent, and the corresponding magnetization curve no longer shows a $1/3$ plateau.

\begin{figure}[htbp]
\includegraphics[width=1\textwidth]{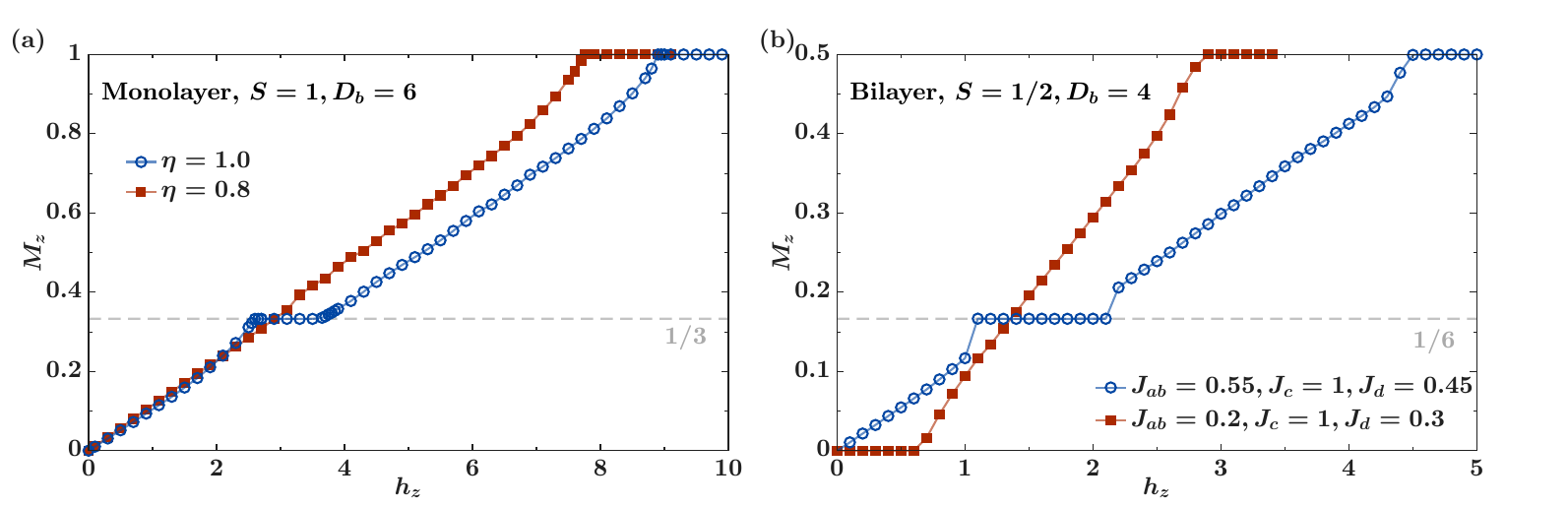}
\caption{
    Magnetization of the representative monolayer and bilayer TLXXZ model. (a) Magnetization of the monolayer TLXXZ model (calculated with $Db_=4$) with $S=1$ for $\eta=1$ (blue circles) and $\eta=0.8$ (red squares). (b) Magnetization of the bilayer system for different coupling strength (calculated with $D_b=4$). The red squares and blue circles correspond to $J_{ab}=0.2 J_c$, $J_d=0.3 J_c$ and $J_{ab}=0.55 J_c$, $J_d=0.45 J_c$, respectively.
}
\label{fig.Mz}
\end{figure}

Similarly, the magnetization of the bilayer system is calculated as the average expectation value of the $z$-components of the six spins, which characterizes multiple plateaus phases in Fig.~4 of the main text. Meanwhile, in addition to the representative process of dimers breaking successively, changing the coupling parameters can induce different magnetization processes. As shown in Fig.~\ref{fig.Mz}(b), when the strengths of $J_d$ and $J_{ab}$ differ substantially, the system no longer exhibits successively breaking of dimers. Instead, once $h_z$ becomes sufficiently strong, it directly forms a canted ferrimagnetic order, in which the spins within each layer are aligned with one another, while the spins in different layers point in different directions. As $h_z$ increases further, the magnitude of the spin expectation value gradually increases, and the spins are polarized toward saturation symmetrically about the field direction, without showing additional plateau phases. This corresponds to a process in which the system evolves directly from the dimer state to the FM state through the BEC state. When $J_{ab}$ and $J_d$ are sufficiently strong, the bilayer system recovers monolayer-like behavior. For example, during the polarization process, the two spins in a spin pair become fully aligned, and the system undergoes a Y-type to UUD-type to V-type polarization process before reaching saturation.

\section{(4) Linear spin wave theory of monolayer TLXXZ model}

Linear spin-wave (LSW) theory provides a qualitative understanding of the main evolutionary trends in the ground-state phase diagram of the monolayer TLXXZ model~\cite{SColpa1978}. In this section, we formulate the LSW theory for the anisotropic monolayer TLXXZ model.

{
LSW theory is built upon the classical ground-state configurations of different phases. By setting $J=S=1$, the classical ground-state energies per spin of the umbrella, $\pi$-coplanar, Y, UUD, and FM phases are respectively given by}
\begin{align}
    \label{Eq.Ecls}
    \notag
    &E_\text{cl, Umbrella}^0 = -\frac{9+h_z^2+18\eta}{6+12\eta}, \\
    \notag
    &E_\text{cl, $\pi$}^0 = -\frac{h_z^2-3h_z(\eta-1)+9(\eta^2+\eta+1)}{9(\eta+1)}, \\
    &E_\text{cl, Y}^0 = -\frac{h_z^2+3h_z(\eta-1)+9(\eta^2+\eta+1)}{9(\eta+1)}, \\
    \notag
    &E_\text{cl, UUD}^0 = -\frac{hz}{3} - \eta, \\
    \notag
    &E_\text{cl, FM}^0 = -h_z + 3\eta.
\end{align}
{Additionally, the classical energy of the V phase can be parameterized by the angle $\theta$ between the magnetic field direction and $S_\text{A}$:}
\begin{align}
    E_\text{cl, V} = \eta
-\frac{1}{4}(3+\eta)\sin^2\theta
-\frac{h_z}{3}\sqrt{4-\sin^2\theta}
\nonumber
+\cos\theta
\left(
-\frac{h_z}{3}
+\eta\sqrt{4-\sin^2\theta}
\right).
\end{align}
{The ground-state energy of the V phase can then be obtained by numerically minimizing $E_\text{cl, V}$ with respect to $\theta$.}

\subsection{{A) Linear spin-wave formalism}}

\begin{figure}[htbp]
\includegraphics[width=0.4\textwidth]{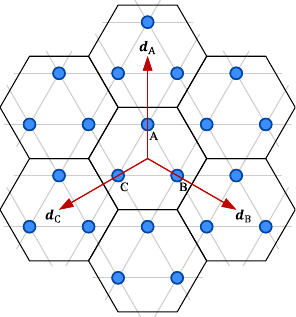}
\caption{
    The unit cell of the monolayer TLXXZ model used in the LSW theory. The black hexagon denotes the unit cell, which contains three sites A, B, and C, and the red arrows indicate the relative displacements of neighboring unit cells.
}
\label{fig.LSW.UnitCell}
\end{figure}

The Hamiltonian in Eq.~(1) of the main text can be recast in a matrix form as
\begin{align}
\matr{H} = \sum_{\braket{mi,nj}_{ab}} \vect{S}_{mi}^{T}\matr{J}_{mi,nj}\vect{S}_{nj} - h_z \sum_{i} S_i^z,
\label{eq.LSW.H}
\end{align}
where $m$ and $n$ label the unit cells located at the real-space positions $\vect{r}_m$ and $\vect{r}_n$, respectively, and the total number of unit cells is denoted by $N=900$. The sublattice indices $i,j\in\pac{{\rm A},{\rm B},{\rm C}}$ refer to the three sites within each unit cell. The spin operators in Eq.~(1) are reorganized into $3\times1$ column vectors, $\vect{S}_{nj}=(S_{nj}^x, S_{nj}^y, S_{nj}^z)^T$, and $\matr{J}_{mi,nj}$ is a $3\times3$ exchange-coupling matrix. As illustrated in Fig.~\ref{fig.LSW.UnitCell}, the unit cell is chosen as a $\vect{Q}=0$ magnetic supercell to accommodate the multi-$\vect{Q}$ magnetic structures of the ground-state configurations. With this choice of unit cell, translational symmetry allows the dependence of the matrix $\matr{J}_{mi,nj}$ on the indices $m$ and $n$ to be reduced to its dependence on the relative displacement $\vect{d}=\vect{r}_n-\vect{r}_m$, namely $\matr{J}_{mi,nj}=\matr{J}_{ij}(\vect{d})$. For nearest-neighbor interactions, the displacement $\vect{d}$ belongs to a set $\vect{d} \in \pac{0, \pm\vect{d}_{\rm A}, \pm\vect{d}_{\rm B}, \pm\vect{d}_{\rm C}}$. When $\pac{i,j,\vect{d}}$ satisfies the nearest-neighbor condition on the triangular lattice, $\matr{J}_{ij}(\vect{d})$ is the diagonal matrix $\matr{J}_0 = {\rm diag}(1,1,\eta)$, with $J_{ab}=1$; otherwise, $\matr{J}_{ij}(\vect{d})$ reduces to the zero matrix.

To begin with, the spin operators $\vect{S}_{nj}$ in Eq.~(\ref{eq.LSW.H}) reduce to classical vectors corresponding to the ground-state configurations of the classical model. The vector $\vect{S}_{nj}$ can be obtained from a classical ferromagnetic order $\vect{S}'_j$ in the local spherical coordinate system through a rotation $\matr{R}_j$:
\begin{align}
    \vect{S}_j = \matr{R}_j \vect{S}'_j.
    \label{eq.LSW.rot}
\end{align}
Taking the direction of the ferromagnetic order as the positive polar axis, the rotation matrix can be written as
\begin{align}
	\label{eq.LSW.R}
\matr{R}_j = 
\begin{pmatrix}
    \cos\theta_j \cos\phi_j & -\sin\phi_j & \sin\theta_j \cos\phi_j \\
    \cos\theta_j \sin\phi_j & \cos\phi_j & \sin\theta_j \sin\phi_j \\
    -\sin\theta_j & 0 & \cos\theta_j
\end{pmatrix},
\end{align}
where $\theta_j$ and $\phi_j$ are the polar and azimuthal angles of $\vect{S}_j$ in the local spherical coordinate system, respectively. To simplify the notation, we further define the complex vector $\vect{u}_j$ from the first two columns of $\matr{R}_j$ and denote the third column of $\matr{R}_j$ by the column vector $\vect{v}_j$. Their components are given by
\begin{align}
	\label{eq.uj.vj}
	u_j^\mu &= R^{\mu 1}_j + \im R^{\mu 2}_j, \\
	v_j^\mu &= R^{\mu 3}_j, \nonumber
\end{align}
where $\vect{v}_j$ is also a unit vector parallel to the classical spin $\vect{S}_j$.

Each ground-state configuration ansatz, or equivalently $\matr{R}_j$, can be constructed by corresponding combinations of $\theta_j$ and $\phi_j$. For example, the classical configuration of the {$\pi$-coplanar} phase can be described as
\begin{align}
    \label{eq.LSW.Ansatz.Fan}
	(\theta_{\rm A},\phi_{\rm A},\theta_{\rm B},\phi_{\rm B},\theta_{\rm C},\phi_{\rm C})_\text{{$\pi$}} 
	&= (0,0,\alpha,0,\alpha,\pi), 
\end{align}
where $\alpha$ is the angle between the spin and the longitudinal field. Similarly, one can construct a classical configuration ansatz for the V phase. Since the transverse magnetization of the V phase is very small, we set it to zero for theoretical simplicity. In this way, the V ansatz can also be parameterized by a single angle $\alpha$ between $S_{\rm A}$ and the longitudinal field:
\begin{align}
    \label{eq.LSW.Ansatz.V}
	(\theta_{\rm A},\phi_{\rm A},\theta_{\rm B},\phi_{\rm B},\theta_{\rm C},\phi_{\rm C})_{\rm V} 
	&= (\alpha,\pi,\arcsin(\frac{1}{2}\sin \alpha),0,\arcsin(\frac{1}{2}\sin \alpha),0).
\end{align}

After the classical spin-configuration ansatz $\matr{R}_j$ is constructed, the three components of the ferromagnetic order $\vect{S}'_j$ are promoted to operators via Holstein-Primakoff transformation as
\begin{align}
\label{eq.LSW.bbdag}
S'^x_{j} &= \frac{\sqrt{2S}}{2} ( b_{j} + b_{j}^\dagger), \\\nonumber
S'^y_{j} &= \frac{\sqrt{2S}}{2\im} ( b_{j} - b_{j}^\dagger), \\\nonumber
S'^z_{j} &= S_j - b_{j}^\dagger b_{j},
\end{align}
where $b_{j}$ and $b_{j}^\dagger$ are the bosonic annihilation and creation operators, respectively. They can be further expressed as a summation over the first Brillouin zone:
\begin{equation}
\label{eq.LSW.bk}
 b_{j} = \frac{1}{\sqrt{N}} \sum_{\vect{k} \in \textrm{B.Z.}} b_{j,\vect{k}} e^{\im\vect{k} \cdot \vect{r}_n}.
\end{equation}

Substituting these into Eq.~(\ref{eq.LSW.H}), we obtain the Hamiltonian of the monolayer TLXXZ model as:
\begin{align}
    \label{eq.LSW.Hk.TLXXZ}
    \matr{H} = S^2 \left\{\frac{1}{S}\sum_{\vect{k} \in \textrm{B.Z.}} \left[ \vect{x}_\vect{k}^\dagger \matr{h}_\vect{k} \vect{x}_\vect{k} - {\rm Tr}\left(\matr{A}_\vect{k}\right) \right] + 3L E_{\rm cl}\right\}.
\end{align}
Here, $\vect{x}_\vect{k}$ is a column vector of bosonic operators corresponding to the three linear spin-wave branches within each unit cell:
\begin{equation}
\vect{x}_\vect{k} = \left[ b_{\vect{k}}^{\rm{A}}, b_{\vect{k}}^{\rm{B}}, b_{\vect{k}}^{\rm{C}}, b_{\vect{-k}}^{\rm{A} \dagger}, b_{\vect{-k}}^{\rm{B} \dagger}, b_{\vect{-k}}^{\rm{C} \dagger} \right]^\intercal.
\end{equation}
The Hermitian matrix $\matr{h}_\vect{k}$ consists of four submatrices:
\begin{align}
	\matr{h}_\vect{k} = 
	\begin{pmatrix}
		\matr{A}_\vect{k} & \matr{B}_\vect{k} \\
		\matr{B}_\vect{k}^\dagger & \overline{\matr{A}}_\vect{-k}
	\end{pmatrix}.
\end{align}
The corresponding submatrices are given by
\begin{align}
	\matr{A}_\vect{k} = 
	\begin{pmatrix}
		\Gamma_{\rm A} & \gamma_\vect{k}^{\rm AB}\Delta_{-}^{\rm AB} & \gamma_\vect{k}^{\rm AC}\Delta_{-}^{\rm AC} \\
		\gamma_\vect{k}^{\rm BA}\Delta_{-}^{\rm BA} & \Gamma_{\rm B} & \gamma_\vect{k}^{\rm BC}\Delta_{-}^{\rm BC} \\
		\gamma_\vect{k}^{\rm CA}\Delta_{-}^{\rm CA} & \gamma_\vect{k}^{\rm CB}\Delta_{-}^{\rm CB}  & \Gamma_{\rm C}
	\end{pmatrix},
\end{align}
and
\begin{align}
	\matr{B}_\vect{k} = 
	\begin{pmatrix}
		0 & \gamma_\vect{k}^{\rm AB}\Delta_{+}^{\rm AB} & \gamma_\vect{k}^{\rm AC}\Delta_{+}^{\rm AC} \\
		\gamma_\vect{k}^{\rm BA}\Delta_{+}^{\rm BA} & 0 & \gamma_\vect{k}^{\rm BC}\Delta_{+}^{\rm BC} \\
		\gamma_\vect{k}^{\rm CA}\Delta_{+}^{\rm CA} & \gamma_\vect{k}^{\rm CB}\Delta_{+}^{\rm CB}  & 0
	\end{pmatrix}.
\end{align}
Defining the reduced external field as $\vect{F}=\pas{0, 0, h_z/S}^T$ and using the compact notation for the rotation-matrix elements in Eq.~(\ref{eq.uj.vj}), the matrix elements of $\matr{h}_\vect{k}$ can be written as
\begin{align}
	\gamma_\vect{k}^{ij} &= 1 + e^{\im \vect{k}\cdot \vect{d}_i} + e^{- \im \vect{k}\cdot \vect{d}_j}, \\
	\Gamma_j &= -\frac{3}{2} \left[\vect{v}_j^T \matr{J}_0 (\vect{v}_i + \vect{v}_l) \right] + \frac{1}{2} \vect{F}^T \vect{v}_j , (i \ne j \ne l), \\
	\Delta_-^{ij} &= \frac{1}{4} \vect{u}_i^T \matr{J}_0 \overline{\vect{u}}_j, \\
	\Delta_+^{ij} &= \frac{1}{4} \vect{u}_i^T \matr{J}_0 \vect{u}_j.
\end{align}
Furthermore, $E_{\rm cl}$ is the reduced classical ground-state energy:
\begin{equation}
	E_{\rm cl} = \displaystyle \sum_{\substack{i,j={\rm A, B, C} \\ i\ne j}} \left(\vect{v}_i^T \matr{J}_0 \vect{v}_j\right) - \frac{\vect{F}^T}{3} \sum_{i={\rm A, B, C}} \vect{v}_i.
\end{equation}
{The classical ground-state energies Eq.~(\ref{Eq.Ecls}) can then be reproduced from $E_{\rm cl}$ with corresponding configuration ansatz.}

The trace of $\matr{A}_{\vect{k}}$, which enters the linear spin-wave correction, is independent of the wave vector $\vect{k}$:
\begin{equation}
	{\rm Tr}\left(\matr{A}_{\vect{k}}\right) = \displaystyle -3 \sum_{\substack{i,j={\rm A, B, C} \\ i\ne j}} \left(\vect{v}_i^T \matr{J}_0 \vect{v}_j\right) + \frac{\vect{F}^T}{2} \sum_{i={\rm A, B, C}} \vect{v}_i.
\end{equation}

Next, one need to determine the spectrum of $\matr{h}_\vect{k}$. An efficient way is to compute the eigenvalues of the matrix $\matr{g}\matr{h}_\vect{k}$, which directly yields the spin-wave spectrum:
\begin{equation}
	\vect{\tilde{\omega}} = (\pm \omega_\vect{k}^{\rm A}, \pm \omega_\vect{k}^{\rm B}, \pm \omega_\vect{k}^{\rm C}).
\end{equation}
Here, the matrix $\matr{g}$ is composed of two $3 \times 3$ identity matrices $\id$:
\begin{equation}
	\matr{g} = 
	\begin{pmatrix}
		\id & 0 \\ 0 & -\id
	\end{pmatrix}.
\end{equation}
Notably, all components of $\vect{\tilde{\omega}}$ should be real, and these components form three pairs, with the two elements in each pair having the same absolute value and opposite signs.

For a given ansatz, the average energy per spin with linear spin wave correction is given by
\begin{equation}
	\label{eq.E_LSW}
	E_{\rm LSW} = S^2 \left( E_{\rm cl} + \frac{1}{S} \delta E \right),
\end{equation}
where the spin wave correction is
\begin{equation}
	\delta E = \frac{1}{3} \pas{- {\rm Tr}\left(\matr{A}_\vect{k} \right) + \frac{1}{L} \sum_{\substack{i={\rm A, B, C} \\ \vect{k}\in{\rm B. Z.}}} \omega_\vect{k}^i }.
\end{equation}

It is also important to emphasize the dependence of both $E_{\rm cl}$ and $E_{\rm LSW}$ on the ansatz angle $\alpha$. At the same point $\pap{\eta,h_z/S}$ in the phase diagram, the angle that minimizes $E_{\rm cl}$ and gives the minimum value $E_{\rm cl}^0$ is not necessarily identical to the angle that minimizes $E_{\rm LSW}$ and gives the minimum value $E_{\rm LSW}^0$. Moreover, take the {$\pi$-coplanar} and V phases as examples, the linear spin-wave energy $E_{\rm LSW}^\text{{$\pi$}}(\alpha)$ obtained from the {$\pi$-coplanar} ansatz and $E_{\rm LSW}^{\rm V}(\alpha)$ obtained from the V ansatz generally have different dependences on $\alpha$. By variationally optimizing $\alpha$, one can determine the optimal angle $\alpha_0$ that minimizes $E_{\rm LSW}$ at a given sampling point in the phase diagram. This yields the lowest energies of the {$\pi$-coplanar} and V phases at that point, denoted by $E_{\rm LSW}^\text{{$\pi$}}(\alpha_0)$ and $E_{\rm LSW}^{\rm V}(\alpha_0)$, respectively. Comparing the relative magnitudes of $E_{\rm LSW}^\text{{$\pi$}}(\alpha_0)$ and $E_{\rm LSW}^{\rm V}(\alpha_0)$ then determines which of the two competing phases is the ground state, and hence gives the phase boundary.

In addition, the LSW theory based on the positive-definiteness assumption~\cite{SColpa1978} makes the UUD and Y phases become energetically higher than the V and {$\pi$-coplanar} phases for $\eta<1$. A more accurate mean-field calculation would require the inclusion of contributions beyond the LSW approximation.

\begin{figure}[htbp]
    \includegraphics[width=0.8\textwidth]{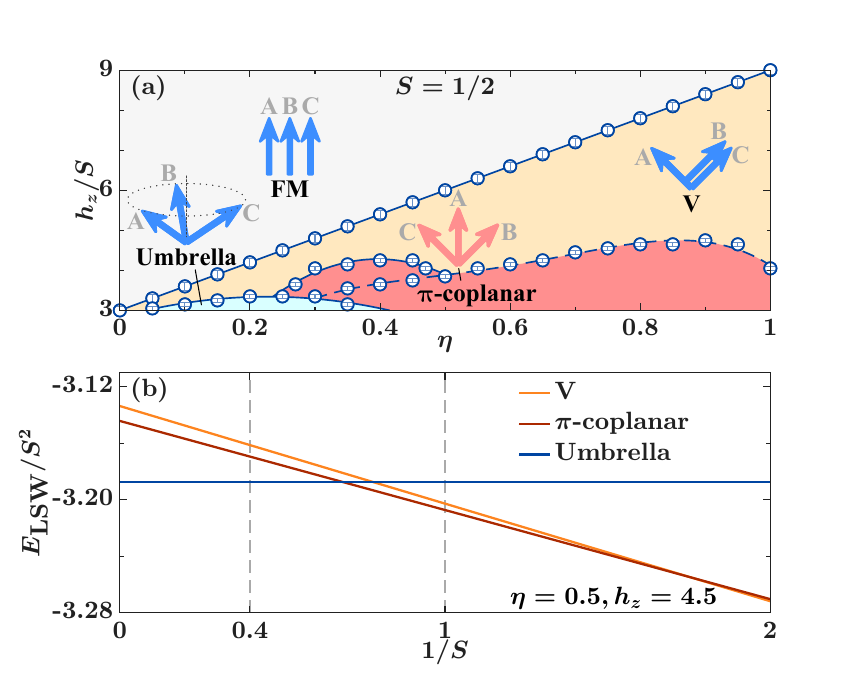}%
    \caption{
    {
        LSW formalism for the existence of $\pi$-coplanar phase. (a) The phase diagram of the monolayer TLXXZ model, obtained via LSW for spin values $S=1/2$. The notation for the phases follows that used in Fig.~A1. 
        (b) LSW Energies of the Fan, V, and umbrella phases as a function of $1/S$ (at $\eta=$ 0.5, $h_z/S=4.5$).
        }
    }
    \label{fig:LSW.slope}
\end{figure}

\subsection{{B) Emergence of the $\pi$-coplanar phase within linear spin-wave theory}}

{
As discussed in the End Matter, although LSW theory predicts a $\pi$-coplanar phase at $S=1/2$ [Fig. \ref{fig:LSW.slope}(a)], the parameter region where it appears does not contradict our rule-of-thumb of parity. 
Furthermore, including higher-order quantum corrections, such as magnon–magnon interactions, or introducing extra variational flexibility, for instance, an adjustable angular parameter for the V phase that removes the constraint of zero transverse magnetization, could further improve the energies of the competing phases and bring the mean-field results into better agreement with the tensor-network data.
} 

{At larger $S$, the LSW correction provides a qualitative picture of how the $\pi$-coplanar phase emerges in a specific region of the phase diagram. In the classical limit ($1/S\to0$), for $\eta<1$, the energies are ordered from lowest to highest as umbrella, $\pi$-coplanar, and V. These classical energies correspond to the intercepts of $E_{\rm LSW}/S^2$ at $1/S=0$. Meanwhile, the spin-wave corrections $\delta E$ ($\leq0$) differ among the three phases, giving different slopes of $E_{\rm LSW}/S^2$ as a function of $1/S$. In contrast to the ordering of the intercepts, for $\eta<1$, the corrections $\delta E$ are ordered from lowest to highest as V, $\pi$-coplanar, and umbrella. The umbrella phase has no spin-wave correction under the positive-definiteness constraint, and hence its $\delta E$ is taken to be zero.}

{The opposite orderings of the intercepts and slopes can lead to an intermediate regime in which the $\pi$-coplanar phase becomes energetically favored: close to the classical limit, the umbrella phase has the lowest energy; for small $S$, the V phase has the lowest energy; while for intermediate $S$, the $\pi$-coplanar phase can become the ground state. Taking the intermediate $\eta=0.5$ as an example [Fig.~\ref{fig:LSW.slope}(b)], the different $1/S$ corrections of the umbrella, $\pi$-coplanar and V phases make the $\pi$-coplanar phase energetically lowest for $S=1$. As $S\rightarrow\infty$, this $\pi$-coplanar region increasingly approaches the classical limit near $\eta=1$.}

\section{Supplementary References}

\end{document}